\documentclass[reprint,amsmath,amssymb,aps]{revtex4-2}

\usepackage{graphicx}
\usepackage{xcolor}

\colorlet{BLUE}{blue}
\colorlet{RED}{red}
\usepackage[ruled,linesnumbered]{algorithm2e}
\usepackage{float}
\usepackage{tikz}
\tikzset{
  gline/.style={line width=1.1pt},
  philine/.style={line width=1.1pt, densely dashed},
  vdot/.style={circle, fill, inner sep=1.1pt}
}
\SetKwInOut{KwInit}{Initialise}

\begin{document}

\title{Spectral theory of large data with power-law latent features}
\author{Philipp Fleig}
\email{philipp.fleig@mr.mpg.de}
\affiliation{Department of Cellular Biophysics, Max Planck Institute for Medical Research,\\
Jahnstra\ss e 29, 69120 Heidelberg, Germany}
\date{\today}

\begin{abstract}
Power-law eigenvalue spectra of covariance matrices estimated from data appear in diverse domains ranging from neural populations to proteins. We introduce a generative latent feature model for such data, optionally corrupted by measurement noise, and develop its spectral theory.
We derive closed leading-order self-consistent equations for the Stieltjes transform of the covariance matrix from which the bulk eigenvalue density is obtained and provide a variational principle for the self-consistent equations.
We also compute the positions and fluctuations of isolated top eigenvalues. The theory is validated in known algebraic limits and tested against simulations.
\end{abstract}

\maketitle

\section{Introduction}
Empirical covariance matrices in many domains have power-law eigenvalue spectra~\cite{stringer_2019, bradde_bialek_2017, qin_colwell_2018, field_1987}. We introduce a probabilistic latent feature model for such data: $N$ observed features (e.g. neurons, pixels, network activations, protein residues, ...) are modelled as linear combinations of $\ell$ hierarchical latent features whose variances decay as a power law $m^{-\alpha}$ of the latent feature index $m$. The observed features are sampled $T$ times and optionally corrupted by additive measurement noise. The technical challenge is to develop a quantitative theory of the eigenvalue density of the empirical covariance matrix of this model as a function of the model parameters, including for the outliers and their fluctuations. The natural framework for treating this problem is random matrix theory (RMT).

In RMT and related fields, power-law spectra have received attention from multiple directions.
In~\cite{qin_colwell_2018}, the Mar\v cenko--Pastur self-consistent relation was applied to power-law population spectra to 
separate phylogeny-induced structure from interactions in protein sequence alignments.
In machine learning, power-law data-covariance spectra underlie neural scaling laws~\cite{bahri_explaining_2024}, and heavy-tailed spectra also appear in the weight matrices of trained networks~\cite{martin_mahoney_2021}. Random-feature and regression models with power-law population spectra have been analysed with free-probability and deterministic-equivalent methods, at the level of bulk spectra and the loss curves they imply~\cite{maloney_roberts_sully_2022, atanasov_scaling_2024, paquette_phases_2024}. At the population level the power-law exponent has been shown to survive a nonlinear random-feature map~\cite{paquette_xiao_zhu_2026}.

In this work we study the sample covariance of the data matrix generated by latent features. In contrast to the models above, its population spectrum is itself random, and its leading eigenvalues grow with the dimension for every $\alpha > 0$.
From an exact representation of the generating function of the eigenvalue density we derive the bulk density together with the hierarchy of isolated top eigenvalues above it, with its position and its fluctuations.
The spectral theory we develop accurately predicts the spectrum controlled by the latent hierarchy, sampling, and isotropic noise alone.
A measured spectrum that departs from this prediction beyond its fluctuation scale carries structure beyond the hierarchy, and the model serves as a reference point from which a richer description can be deemed necessary.
Thus, we propose the model as a generative null model for data whose only structure is a power-law hierarchy of latent variances.

Two model parameters organise the regimes that the theory covers: the power-law exponent $\alpha$ and the latent aspect ratio $\ell/N$. The power-law exponent separates into a super- and a sub-critical regime, above and below $\alpha = 1$, respectively.
The theory we develop covers the proportional scaling regime, where $\ell$, $N$, and $T$ grow at fixed ratios. We also consider both the regime of few latent features, $\ell \ll N$, and the latent-rich regime, $\ell \gg N$, in which the hierarchy of latent features outnumbers the observed features. With these regimes the model covers data with qualitatively different properties.

The article is organised as follows. Section~\ref{sec:setup} defines the model. Section~\ref{sec:exact} derives an exact representation of the generating function, from which Section~\ref{sec:planar} obtains the self-consistent equations for the bulk eigenvalue density, validates them against simulations, and derives a variational principle for them. Section~\ref{sec:continuum} takes the continuum limit in the latent feature index.
Section~\ref{sec:edge} treats the isolated top eigenvalues, computing their positions and fluctuations. Section~\ref{sec:discussion} concludes with a discussion of model limitations and open problems.

\section{Model and conventions}
\label{sec:setup}
\begin{figure}[t]
\includegraphics[width=\columnwidth]{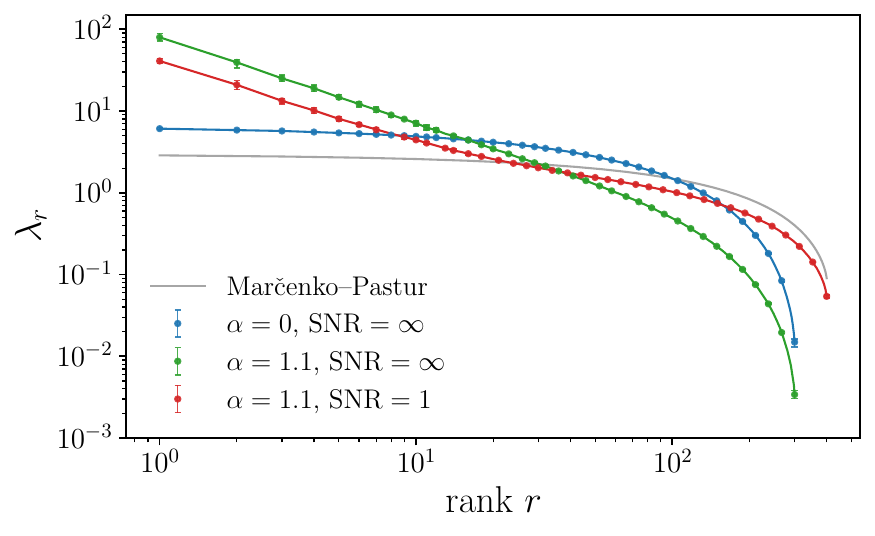}\\
\includegraphics[width=\columnwidth]{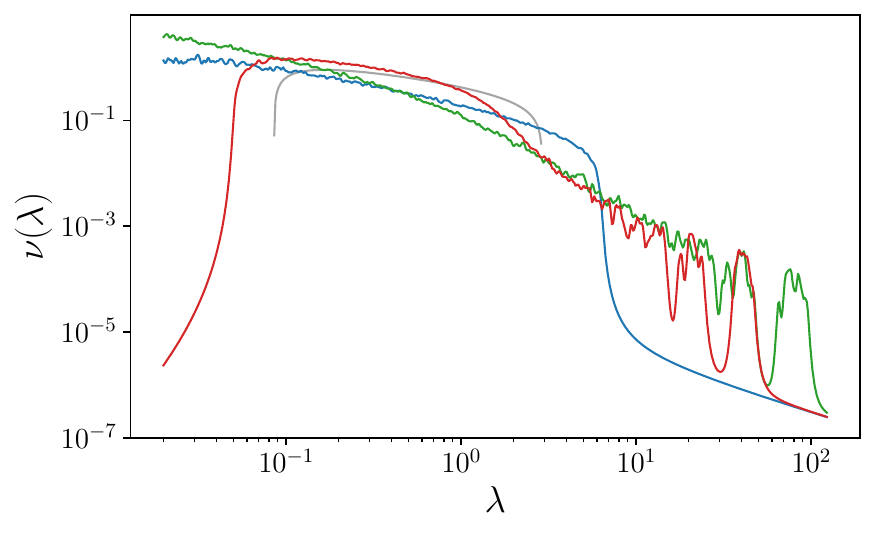}
\caption{Spectral phenomenology of the power-law latent feature model.
Top: ranked eigenvalue spectra. Bottom: the corresponding eigenvalue densities. Blue: $\alpha = 0$, noiseless. Green: $\alpha = 1.1$, noiseless. Red: $\alpha = 1.1$ at $\mathrm{SNR} = 1$. Grey: the Mar\v cenko--Pastur distribution of pure noise at aspect ratio $q$, in the same trace normalisation. Simulations of the model at $N = 400$, $\ell = 300$, and $T = 800$ ($q = 0.5$ and $q_V = 0.75$). Lines are the rank-wise means over 20 independent realisations. Points mark log-spaced ranks. Error bars mark the 10th--90th percentile range over realisations and are small. The $N - \ell$ zero eigenvalues of the noiseless spectra are omitted on the logarithmic axes. The densities are averaged over the same realisations and smoothed at width proportional to $\lambda$, to make the smoothing uniform on the log axes.}
\label{fig:hook-ranks}
\end{figure}

We consider the generative probabilistic latent feature model for the $T\times N$ data matrix $\mathbf{X}$
\begin{align}
\mathbf{X} := \mathbf{U} \mathbf{V} + \sigma \mathbf{R} \in \mathbb{R}^{T\times N}\,,
\end{align}
with mutually independent Gaussian factors, with independent entries,
\begin{gather}
U_{tm} \sim \mathcal{N}(0, 1)\,,\quad
V_{mn} \sim \mathcal{N}(0, m^{-\alpha})\,,\nonumber\\[4pt]
R_{tn} \sim \mathcal{N}(0, 1)\,,
\end{gather}
for features $n = 1, \ldots, N$ recorded over samples $t = 1, \ldots, T$. The model uses the coefficient matrix $\mathbf{U}$ to linearly mix latent features $m = 1, \ldots, \ell$, the rows of the feature matrix $\mathbf{V}$, with variance power-law exponent $\alpha \ge 0$. The matrix $\mathbf{U}$ thus holds the sampled latent feature activities. The model also includes additive intrinsic noise $\sigma\mathbf{R}$ with noise amplitude $\sigma \ge 0$.
The model's three matrix dimensions $N$, $T$, and $\ell$ are independent and we define the aspect ratios
\begin{align}
q := \frac{N}{T}\,,\quad q_V := \frac{\ell}{N}\,,\quad q_U := \frac{\ell}{T} = q\,q_V\,.
\end{align}
We note that for exponent value $\alpha=0$, the model reduces to the latent feature model of~\cite{fleig_nemenman_2022}. 

Our goal is to develop a theory for the eigenvalue spectrum of the empirical covariance matrix of the observations,
\begin{align}
\mathbf{W}_c := \frac{\mathbf{X}_c^T \mathbf{X}_c}{(T - 1)\,\gamma^2}\,,\quad
(\mathbf{X}_c)_{tn} := X_{tn} - \bar X_n\,,
\end{align}
which is an $N\times N$ finite-size random matrix.
In this definition $\bar X_n := T^{-1}\sum_{t} X_{tn}$ are the column means and $\gamma^2$ is a normalisation constant fixed below. Since the coefficient matrix $\mathbf{U}$ and noise matrix $\mathbf{R}$ are Gaussian, and therefore invariant under rotations of the sample index, subtracting the column means is equivalent to discarding one sample: $\mathbf{W}_c$ at $T$ samples is distributed, exactly, as the uncentred covariance matrix of the same model at $T - 1$ samples. Following the treatment of centring in~\cite{bun_bouchaud_potters_2017}, we therefore develop the theory for the normalised Gram matrix of uncentred data
\begin{align}\label{eq:W}
\mathbf{W} := \frac{\mathbf{X}^T \mathbf{X}}{T\,\gamma^2}\,,
\end{align}
at arbitrary sample count $T$; applied at $T - 1$, it covers the centred $\mathbf{W}_c$ exactly. The distribution of $\mathbf{W}$ induced by the Gaussian factors $\mathbf{U}$, $\mathbf{V}$, and $\mathbf{R}$ is the random matrix ensemble studied in this work.

To determine a natural normalisation we compute the mean diagonal entry of the Gram matrix, $\mathbf{X}^T\mathbf{X}$, under the model, $\mathbb{E}[(\mathbf{X}^T\mathbf{X})_{ii}] = T(S_\alpha + \sigma^2)$, where
\begin{align}
    S_\alpha:=\sum_{m=1}^\ell m^{-\alpha}
\end{align}
is the generalised harmonic sum, cf. Appendix~\ref{app:salpha}. The normalisation choice
\begin{align}
\gamma^2 = S_\alpha + \sigma^2
\end{align}
thus fixes the diagonal elements in Eq.~\eqref{eq:W} to $\mathbb{E}[W_{ii}] = 1$, which is the normalisation in which the spectrum is of order one for $\alpha=0$. Matrices are set in boldface. Throughout, $\mathbb{E}$ without subscript averages over all model randomness, and, where distinction matters, a subscript, as in $\mathbb{E}_U$, restricts the average to the indicated factor.

It will be convenient to introduce separate symbols $s$ and $n_\sigma$ for the signal and noise fractions, respectively,
\begin{align}\label{eq:sn}
s := \frac{S_\alpha}{\gamma^2}\,,\quad n_\sigma := \frac{\sigma^2}{\gamma^2}\,,\quad s + n_\sigma = 1\,,
\end{align}
so that the noiseless model has $s = 1$, $n_\sigma = 0$. The same information is carried by the signal-to-noise ratio (SNR)
\begin{gather}
\mathrm{SNR} := \frac{S_\alpha}{\sigma^2} = \frac{s}{n_\sigma}\,,\nonumber\\
\label{eq:snr-def}
{s} = \frac{\mathrm{SNR}}{1 + \mathrm{SNR}}\,,\quad
n_\sigma = \frac{1}{1 + \mathrm{SNR}}\,;
\end{gather}
the total latent variance received by each observed variable, $\mathbb{E}[(\mathbf{U}\mathbf{V})_{tn}^2] = S_\alpha$, relative to its noise variance $\sigma^2$.
The noiseless model is the limit $\mathrm{SNR} \to \infty$ and pure noise is $\mathrm{SNR} \to 0$.

Figure~\ref{fig:hook-ranks} illustrates the model's spectral phenomenology which the theory must quantify. We show the simulated data for different parameter values in two representations: the eigenvalues $\lambda_r$ of the normalised Gram matrix $\mathbf{W}$, Eq.~\eqref{eq:W}, ranked by decreasing magnitude (top), together with the corresponding eigenvalue densities (bottom). For reference we show the Mar\v cenko--Pastur distribution for pure noise data (grey). At $\alpha = 0$ without measurement noise (infinite SNR), the ranked spectrum is flat and of order one, showing pure sampling structure with no latent hierarchy (the case of~\cite{fleig_nemenman_2022}). At $\alpha = 1.1$ and infinite SNR, the leading latent channels stand out as isolated outlier eigenvalues above the bulk, a discrete top on a power law. At finite measurement noise (finite SNR) the top of the spectrum is cut off from below at the noise floor. The densities provide a complementary perspective: at $\alpha = 0$ an order-one bulk with no structure beyond its edges, at $\alpha = 1.1$ a power-law bulk carrying the outliers at its top, and at finite noise a noise bulk rising from below while the surviving outliers are pulled inward. The quantitative theory that we develop computes each of these spectral characteristics.

\subsection{Calculation setup}
The resolvent of $\mathbf{W}$ is the matrix $(z \mathbf{I}_N - \mathbf{W})^{-1}$. The starting point is its normalised trace, the empirical Stieltjes transform,
\begin{align}
\hat g_N(z) := \frac{1}{N}\,\mathrm{tr}\,(z \mathbf{I}_N - \mathbf{W})^{-1}\,.
\end{align}
It is a function of the random realisation of $\mathbf{W}$ and the spectral parameter, $z=\lambda+i\eta$, with $\eta>0$. Its boundary values give the eigenvalue density of the data, $\hat\nu_N(\lambda) = -\pi^{-1}\mathrm{Im}\,\hat g_N(\lambda + i0^+)$, by the Sokhotski--Plemelj formula.
Proportional growth of the model's three matrix dimensions at fixed ratios $q$ and $q_V$ is called the \emph{proportional scaling}. As the dimensions grow in the proportional scaling, the empirical Stieltjes transform \emph{self-averages}: for almost every realisation it differs from a deterministic function $g(z)$ by an amount that vanishes with increasing $N$~\cite{marchenko_pastur_1967, silverstein_bai_1995}. In standard terminology, $g$ is the \emph{deterministic equivalent} of the empirical transform~\cite{hachem_loubaton_najim_2007}.
Although the self-averaging statement is asymptotic, the deterministic equivalent is defined at the actual finite dimensions $N$, $T$, and $\ell$, and we evaluate it there rather than in an $N \to \infty$ limit.
By rotational invariance of the observation coordinates, the deterministic equivalent of the resolvent matrix itself is $g(z)\,\mathbf{I}_N$, so no information is lost in passing to the scalar. Its boundary values give the leading-order density $\nu_0(\lambda) := -\pi^{-1}\mathrm{Im}\,g(\lambda + i0^+)$.

The deterministic equivalent $g$ is the target of the bulk theory. It describes the spectrum where the eigenvalues lie dense and the density is the meaningful description. The isolated top eigenvalues (outliers) lie outside this description. Each is a single eigenvalue, with spectral mass $1/N$, whose position scatters between realisations at the relative scale $N^{-1/2}$. We compute the positions and the fluctuation law of the outliers separately, building on the bulk solution. The key object of the calculation is the finite-$N$ mean Stieltjes transform,
\begin{align}
g_N(z) := \mathbb{E}[\hat g_N(z)]\,,
\end{align}
with the expectation taken over the three Gaussian factors $\mathbf{U}$, $\mathbf{V}$, and $\mathbf{R}$, whose boundary values give the mean eigenvalue density $\nu_N(\lambda) = -\pi^{-1}\mathrm{Im}\,g_N(\lambda + i0^+)$. At leading order the mean transform coincides with the deterministic equivalent, $g_N = g + O(1/N)$, and it is through $g_N$ that we compute $g$.

For the large-$N$ analysis of $g_N$ we start from the observation that the mean transform is a $z$-derivative of a log-determinant, $g_N = N^{-1}\partial_z\,\mathbb{E}[\mathrm{tr}\log(z \mathbf{I}_N - \mathbf{W})]$ (we use its trace form throughout). This expectation of a logarithm is known as the quenched average~\cite{edwards_jones_1976}. Its direct evaluation is hard, so we compute instead the annealed generating function~\cite{livan_novaes_vivo_2018}
\begin{align}
\label{eq:Z-def}
Z(z) := \mathbb{E}\bigl[\det(z \mathbf{I}_N - \mathbf{W})^{-1/2}\bigr]\,,
\end{align}
with the same expectation as in $g_N$. The annealed average replaces the average of the logarithm by the logarithm of the average, and is evaluated below by reducing it to a one-dimensional integral and applying the saddle-point (Laplace) method. The generating function fixes the mean transform up to $O(1/N)$,
\begin{align}\label{eq:mean_logZ}
g_N = -\frac2N\partial_z\log Z + O(1/N)\,.
\end{align}
The annealed and quenched averages agree at leading order and differ at $O(1/N)$, which is beyond the order of this work. Throughout the article we work at physics-level rigour: the algebra is exact, and the asymptotic steps are controlled by explicit numerical checks.

\section{Exact radial representation}
\label{sec:exact}

We reduce the generating function in Eq.~\eqref{eq:Z-def}, at finite size $N$, to an exact one-dimensional integral. Every subsequent approximation is applied to this integral, with an explicit error term. We start from the representation of the determinant in Eq.~\eqref{eq:Z-def} as a Gaussian integral~\cite{edwards_jones_1976}
\begin{align}
\label{eq:det-rep}
\det(z\mathbf{I}_N - \mathbf{W})^{-1/2} = (2\pi)^{-N/2}\int d\psi\;e^{-\frac{1}{2}\psi^T(z\mathbf{I}_N - \mathbf{W})\psi}\,,
\end{align}
with $\psi \in \mathbb{R}^N$, which is convergent when $z\mathbf{I}_N - \mathbf{W}$ is positive definite, i.e. for real $z$ above the top eigenvalue, and extended off the spectrum by analytic continuation. The representation in Eq.~\eqref{eq:det-rep} linearises the dependence on $\mathbf{W}$ through $\psi^T \mathbf{W}\psi = |\mathbf{X}\psi|^2/(T\gamma^2)$. We make the following useful observation. Conditioned on $\mathbf{U}$ (whose average we defer), and at fixed $\psi$, the vector $\mathbf{X}\psi = \mathbf{U}\mathbf{V}\psi + \sigma \mathbf{R}\psi$ is linear in the Gaussian factors $\mathbf{V}$ and $\mathbf{R}$, and hence itself a Gaussian vector in $\mathbb{R}^T$, with zero mean and covariance
\begin{align}
\mathrm{Cov}_{V,R}(\mathbf{X}\psi) &:= \mathbb{E}_{V,R}\bigl[(\mathbf{X}\psi)(\mathbf{X}\psi)^T\mid \mathbf{U}\bigr]\nonumber\\
&= |\psi|^2\bigl(\mathbf{U}\mathbf{\Sigma}\mathbf{U}^T + \sigma^2 \mathbf{I}_T\bigr)\,,
\end{align}
where $\mathbf{\Sigma} := \mathrm{diag}(m^{-\alpha})_{m=1}^{\ell}$ is the diagonal matrix of latent feature variances.
The cross terms between $\mathbf{U}\mathbf{V}\psi$ and $\sigma \mathbf{R}\psi$ vanish in expectation because $\mathbf{V}$ and $\mathbf{R}$ are independent with zero mean. The average over $\mathbf{V}$ and $\mathbf{R}$, exchanged with the $\psi$-integral of Eq.~\eqref{eq:det-rep} and taken at fixed $\psi$, is therefore a single Gaussian integral over the vector $\mathbf{X}\psi \in \mathbb{R}^T$. Writing the average against the Gaussian density of $\mathbf{X}\psi$ and combining the exponents gives a ratio of Gaussian normalisations,
\begin{align}
\mathbb{E}_{V,R}\bigl[e^{\frac{|\mathbf{X}\psi|^2}{2T\gamma^2}}\bigr]
= \det\!\Bigl(\mathbf{I}_T - \frac{\mathrm{Cov}_{V,R}(\mathbf{X}\psi)}{T\gamma^2}\Bigr)^{-1/2}\,.
\end{align}
The integrand now depends on $\psi$ only through $r := |\psi|^2$, so the $\psi$-integral reduces to its radial part, $\int d\psi\,f(|\psi|^2) = \frac{\pi^{N/2}}{\Gamma(N/2)}\int_0^\infty dr\,r^{N/2-1}f(r)$. Collecting the pieces gives the radial form of the annealed generating function in Eq.~\eqref{eq:Z-def},
\begin{align}
\label{eq:Z-radial}
Z(z) = c_N\int_0^\infty dr\;r^{N/2-1}\,e^{-zr/2}\;F(r)\,,
\end{align}
where $c_N$ is a $z$-independent constant and
\begin{align}
\label{eq:F-def}
F(r) := \mathbb{E}_U\biggl[\det\!\Bigl(\mathbf{I}_T - \frac{r}{T\gamma^2}\bigl(\mathbf{U}\mathbf{\Sigma}\mathbf{U}^T + \sigma^2 \mathbf{I}_T\bigr)\Bigr)^{-1/2}\biggr]\,.
\end{align}
All dependence on the random factors is now carried by the single scalar function $F$.

To prepare the large-$N$ asymptotics, we rewrite the determinant in Eq.~\eqref{eq:F-def}. In the proportional scaling the three model dimensions grow large together, in fixed ratios. We generically refer to this joint growth as large $N$, while for each asymptotic step we name the dimension whose growth controls its error.
We first define rescaled latent channel weights $w_m$ and the rescaled weight matrix $\mathbf{\Sigma}_w$
\begin{align}
\label{eq:w-def}
w_m &:= \frac{\ell}{S_\alpha}\,m^{-\alpha}\,,\quad
\overline{w} = 1\,,\nonumber\\
\mathbf{\Sigma}_w &:= \mathrm{diag}(w_m)\,,\quad \mathbf{\Sigma} = \frac{S_\alpha}{\ell}\,\mathbf{\Sigma}_w\,,
\end{align}
where $\overline{w^k} := \ell^{-1}\sum_{m=1}^{\ell} w_m^k$ denotes the $k$-th moment of the channel weights. The top weight $w_1 = \ell/S_\alpha$ grows with $\ell$ for every $\alpha > 0$, as $\ell/\zeta(\alpha)$ for $\alpha > 1$, as $\ell/\log\ell$ at $\alpha = 1$, and as $(1 - \alpha)\,\ell^{\alpha}$ for $\alpha < 1$, by the asymptotics of $S_\alpha$ in Appendix~\ref{app:salpha}. Only at $\alpha = 0$ are all weights equal to one.
We also define the coloured Wishart matrix~\cite{burda_correlated_2005, silverstein_bai_1995}
\begin{align}
\mathbf{Y} := \frac{\mathbf{U}\,\mathbf{\Sigma}_w\,\mathbf{U}^T}{T} \in \mathbb{R}^{T\times T}\,,
\end{align}
which is a normalised Gram matrix of $T$ i.i.d. samples from $\mathcal{N}(0, \mathbf{\Sigma}_w)$, with population covariance $\mathbf{\Sigma}_w$.

With the change of variables $r = N g$ (the factor $N$ is chosen for convenience so that, at the saddle of the radial integral, $g$ becomes the leading-order Stieltjes transform itself, as the saddle-point analysis below shows) and the signal and noise fractions $s$ and $n_\sigma$, the determinant argument in Eq.~\eqref{eq:F-def} separates into a scalar part and a $\mathbf{Y}$ part:
\begin{align}
\frac{r}{T\gamma^2}\bigl(\mathbf{U}\mathbf{\Sigma}\mathbf{U}^T + \sigma^2 \mathbf{I}_T\bigr)
= b\,\mathbf{I}_T + (1 - b)\,a\,\mathbf{Y}\,,
\end{align}
where
\begin{align}
\label{eq:ab-def}
a(g) := \frac{s\,g}{q_V\,(1 - b)}\,,\quad
b(g) := q\,n_\sigma\,g\,,
\end{align}
obtained using Eqs.~\eqref{eq:sn} and~\eqref{eq:w-def}.
The coefficient functions $a$ and $b$ afford a direct interpretation. The function $b$ is the strength of the noise dressing, whose geometric series produces the screening factors $(1 - b)^{-1}$ of Eq.~\eqref{eq:closure-noisy-raw}. The function $a$ is the noise-dressed coupling at which the signal factor is probed. In other words, it is the bare coupling $g/q_V$, rescaled by the signal fraction and screened by one power of $(1 - b)^{-1}$. From the change of variables and after factoring out $(1 - b)$, we obtain
\begin{align}
\label{eq:F-factorised}
F(Ng) = (1 - b)^{-T/2}\;\mathbb{E}_U\Bigl[\det\bigl(\mathbf{I}_T - a\,\mathbf{Y}\bigr)^{-1/2}\Bigr]\,.
\end{align}
Everything non-trivial is the average of an inverse square root of the characteristic polynomial of $\mathbf{Y}$.

\section{Self-consistent equations at leading order}
\label{sec:planar}

In this section we derive the self-consistent equations for the deterministic equivalent $g$, obtain a variational principle for those equations, and make a direct comparison of the theory with simulations.

\subsection{Saddle of the radial integral}
Starting from Eq.~\eqref{eq:Z-radial} together with Eq.~\eqref{eq:F-factorised} we rewrite the exact radial representation in a form suited for application of Laplace's method:
\begin{align}
\label{eq:Z-laplace}
Z(z) \propto \int_0^\infty dg\,\exp\Bigl(-\frac{N}{2}\,\mathcal{F}(g) + O(1)\Bigr)\,,
\end{align}
with the free energy
\begin{align}
\label{eq:free-energy}
\mathcal{F}(g) = z g - \log g + \frac{1}{q}\log(1 - b) + q_V\,\omega(a)\,,
\end{align}
where $a$ and $b$ are the functions in Eq.~\eqref{eq:ab-def} and
\begin{align}
\label{eq:omega-def}
\omega(x) := \frac{1}{\ell}\,\mathbb{E}_U\Bigl[\mathrm{tr}\log\bigl(\mathbf{I}_T - x \mathbf{Y}\bigr)\Bigr]
\end{align}
is the expected log-determinant of the coloured Wishart matrix $\mathbf{Y}$ per channel, evaluated at leading order. The term arises from passing the expectation over $\mathbf{U}$ through the exponential, replacing the random exponent $\mathrm{tr}\log(\mathbf{I}_T - x\,\mathbf{Y})$ by its mean. This is justified since deformations of the spectrum are suppressed more strongly than the exponential favours them, and the average is dominated by typical realisations~\cite{benarous_guionnet_1997, hiai_petz_1998}. The step gave the term $q_V\,\omega(a)$ of the free energy, at the cost of the $O(1)$ contribution in Eq.~\eqref{eq:Z-laplace}, which does not affect the leading order.

At large $N$ we apply Laplace's method to Eq.~\eqref{eq:Z-laplace} to obtain, up to a $z$-independent constant,
\begin{align}
\log Z = -\frac{N}{2}\mathcal{F}(g_*) + O(1)\,,
\end{align}
where $g_*$ is the saddle point defined by $\mathcal{F}'(g_*) = 0$ with prime denoting $d/dg$ at fixed $z$ (throughout, a prime denotes a derivative with respect to the function's displayed argument). Thus, $\mathcal{F}$ is the free energy density. On the branch of solutions continued from the free solution $g = 1/z$, the saddle is a local minimum of $\mathcal{F}$. To leading order, the mean Stieltjes transform $g_N$ is given by the saddle point. In Eq.~\eqref{eq:mean_logZ}, the $z$-derivative acts both explicitly and implicitly through $g_*(z)$. The implicit part drops out because $\mathcal{F}'(g_*) = 0$, and the explicit $z$-dependence of $\mathcal{F}$ is the single term $zg$, giving
\begin{align}
\label{eq:saddle-z}
\partial_z\,\mathcal{F}(g_*(z); z) = g_*(z)\,,
\end{align}
so $g_N(z) = g_*(z)$ at leading order. The stationarity condition is therefore a closed equation satisfied by the leading-order transform $g$. The next step is to derive its explicit form.

The saddle equation $\mathcal{F}' = 0$ requires the two derivatives
\begin{align}
\frac{d b}{d g} = q n_\sigma\,,\quad \frac{da}{d g} = \frac{s}{q_V\,(1 - b)^2}\,,
\end{align}
where the numerator simplifies because $b$ is itself proportional to $g$. 
Stationarity gives
\begin{align}
\label{eq:closure-noisy-raw}
z - \frac{1}{g} = \frac{n_\sigma}{1 - b} + \frac{s}{(1 - b)^2}\;\mathcal{K}(a)\,,
\end{align}
where
\begin{align}
    \mathcal{K}(x):=-\omega'(x) = \frac1\ell\mathbb{E}_U\bigl[\mathrm{tr}[\mathbf{Y}(\mathbf{I}_T - x\mathbf{Y})^{-1}]\bigr]\,,
\end{align}
i.e. the trace of the resolvent-type function of $\mathbf{Y}$.
The stationarity condition, Eq.~\eqref{eq:closure-noisy-raw}, is a \emph{Dyson equation}. The free part $zg - \log g$ of $\mathcal F$ alone would place the saddle at the free value $g = 1/z$. The interaction terms enter stationarity only as an additive shift of the spectral parameter, $g = (z - \Sigma(g))^{-1}$, with $\Sigma(g)$ the right-hand side of Eq.~\eqref{eq:closure-noisy-raw}. This identifies $\Sigma$ as the self-energy in the standard sense, which dresses the bare spectral parameter.

The self-energy $\Sigma$ is the sum of a noise term, given by the Mar\v cenko--Pastur self-energy of the noise factor alone, and a signal term, given by the noiseless self-energy weighted by the signal fraction, probed at the dressed coupling $a$ and screened by two powers of $(1 - b)$. This additive-with-dressing structure is the standard form of self-energies for independent additive contributions~\cite{sengupta_mitra_1999}.

\subsection{Deterministic equivalents of the coloured Wishart}

The remaining task is to evaluate $\mathcal{K}(x)$ at leading order. The result is a deterministic equivalent for the resolvent $(\mathbf{I}_T - x\mathbf{Y})^{-1}$ probed along each channel direction $u_m$ (the $m$-th column of $\mathbf{U}$). The leading-order evaluation of $\mathcal{K}$ follows from it. The full derivation is given in Appendix~\ref{app:silverstein}. The result is expressed in terms of the scalar function
\begin{align}
\label{eq:phi-def}
\phi(x) := \frac{1}{T}\,\mathbb{E}_U\Bigl[\mathrm{tr}\bigl(\mathbf{I}_T - x\,\mathbf{Y}\bigr)^{-1}\Bigr]\,,
\end{align}
i.e. the expectation of the normalised trace of the coloured Wishart's resolvent. It satisfies the scalar self-consistency
\begin{align}
\label{eq:phi-eq}
\frac{\phi - 1}{\phi} = q_U\,x\;\frac{1}{\ell}\sum_{m=1}^{\ell}\frac{w_m}{1 - x\,\phi\,w_m}\,,
\end{align}
where the physical solution is selected by $\phi \to 1$ as $x \to 0$. Expressed in terms of $\phi$ we have
\begin{align}
\label{eq:K-eq}
\mathcal{K}(x) = \frac{1}{\ell}\sum_{m=1}^{\ell}\frac{\phi\,w_m}{1 - x\,\phi\,w_m}\,.
\end{align}
Each latent channel is probed at the rescaled coupling $x\phi$ rather than $x$. In free-probability language $\phi$ is also known as the subordination function of the multiplicative convolution~\cite{bun_bouchaud_potters_2017}.

\subsection{The self-consistent equations}

Assembling Eqs.~\eqref{eq:closure-noisy-raw}--\eqref{eq:K-eq} gives the leading-order self-consistent equations of the model at general noise level,
\begin{align}
\label{eq:closure-noisy}
\begin{aligned}
z - \frac{1}{g} &= \frac{n_\sigma}{1 - q\,n_\sigma\,g} + \frac{s}{(1 - q\,n_\sigma\,g)^2}\;\frac{1}{\ell}\sum_{m=1}^{\ell}\frac{\phi\,w_m}{1 - a\,\phi\,w_m}\,,\\[2pt]
\frac{\phi - 1}{\phi} &= \frac{q\,s\,g}{1 - q\,n_\sigma\,g}\;\frac{1}{\ell}\sum_{m=1}^{\ell}\frac{w_m}{1 - a\,\phi\,w_m}\,,
\end{aligned}
\end{align}
where $\phi = \phi(a)$. The second line is the $\phi$ equation Eq.~\eqref{eq:phi-eq} at the dressed coupling $x = a$. In the noiseless case, $s = 1$, $n_\sigma = 0$, $b = 0$, and $a = g/q_V$, the pair of equations specialises to
\begin{align}
\label{eq:closure}
\begin{aligned}
z - \frac{1}{g} &= \frac{1}{\ell}\sum_{m=1}^{\ell}\frac{\phi\,w_m}{1 - (g/q_V)\,\phi\,w_m}\,,\\[2pt]
\frac{\phi - 1}{\phi} &= q\,g\;\frac{1}{\ell}\sum_{m=1}^{\ell}\frac{w_m}{1 - (g/q_V)\,\phi\,w_m}\,.
\end{aligned}
\end{align}

The pair of equations in Eq.~\eqref{eq:closure-noisy} admits an exact reduction to a single equation.
Since $\phi$ is a scalar it factors out of the channel sum, so the second line of Eq.~\eqref{eq:closure-noisy} can be evaluated on the first, giving
\begin{align}
\label{eq:phi-identity-noisy}
\phi(z) = (1 - q\,n_\sigma\,g)\bigl[1 + q\,(z g - 1)\bigr]\,.
\end{align}
This is an exact identity at every $z$ on the solution of the equations. In the noiseless case where $n_\sigma=0$, the right-hand side reduces to the second factor.
Substituting back into the Dyson equation leaves a single implicit scalar equation for $g(z)$.

\subsection{Consistency checks and comparison with simulation}
\label{sec:checks}

\begin{figure*}[t]
\includegraphics[width=\textwidth]{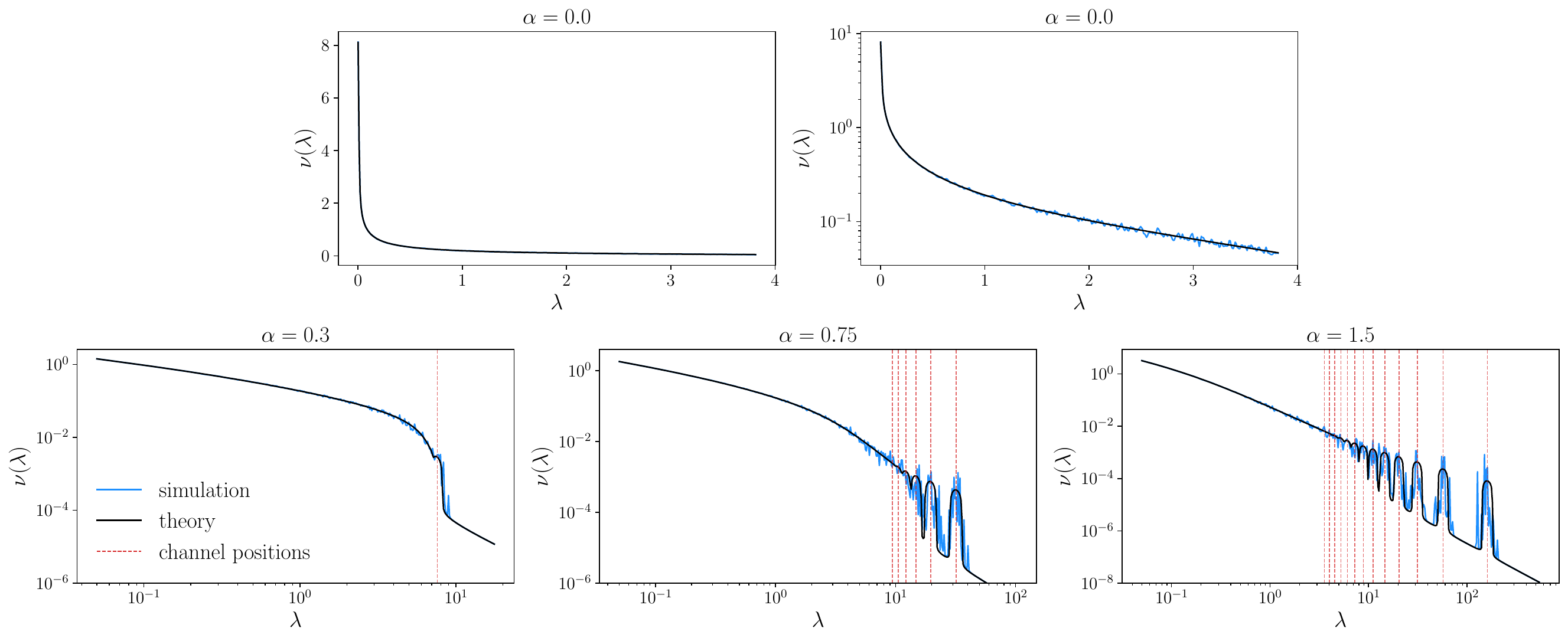}
\caption{Theory versus simulated densities across the power-law exponent regimes. Panels: exponent values $\alpha \in \{0, 0.3, 0.75, 1.5\}$, without noise. The $\alpha = 0$ panel is shown on linear and on logarithmic density axes. Blue: Lorentzian-smoothed empirical density. Black: density from the self-consistent equations. Red dashed vertical lines: channel positions, the roots of the outlier condition, Eq.~\eqref{eq:outlier-noisy}. All panels at $N = 400$, $\ell = 300$, and $T = 800$ ($q = 0.5$ and $q_V = 0.75$). Both densities are smoothed at width $\eta = 0.01$. The empirical density is averaged over 80 independent realisations.}
\label{fig:overlay-alpha}
\end{figure*}

We verify the self-consistent equations by analytic checks of limits and direct comparison with simulations. We start with the exact algebraic checks.

As an immediate limit check, we note that in the pure-noise limit where $s \to 0$, $n_\sigma \to 1$ ($\mathrm{SNR} \to 0$), the signal term in Eq.~\eqref{eq:closure-noisy} drops out since it carries the prefactor $s$, leaving $z - 1/g = 1/(1 - qg)$. This is the standard Mar\v cenko--Pastur (MP) self-consistent equation at aspect ratio $q$ in our normalisation, as it must be.
As a second check, we consider the case $\alpha=0$ which we can compare to the free-probability results of~\cite{fleig_nemenman_2022} for this model.
Specifically, at $\alpha = 0$ the self-consistent equations collapse to a cubic polynomial in the noiseless case, and a quartic at finite noise. Details of the comparison are given in Appendix~\ref{app:alpha0}. Furthermore, for all values $\alpha\geq0$ the equations reproduce the exact second moment, computed by Wick contractions with no input from the equations, cf. Appendix~\ref{app:moments}.

Finally, direct Monte Carlo simulations confirm the self-consistent equations across the exponent regimes. We solve the self-consistent equations numerically by damped fixed-point iteration, cf. Algorithm~\ref{algo:solver} of Appendix~\ref{app:solver}. Figure~\ref{fig:overlay-alpha} shows the density overlays at $N = 400$, $\ell = 300$, $T = 800$ without noise, for $\alpha\in\{0,0.3,0.75,1.5\}$. Theory and empirical densities are both Lorentzian-smoothed at the same width $\eta$, which is the imaginary offset at which the transform is evaluated, $\nu_\eta(\lambda) = -\pi^{-1}\mathrm{Im}\,g(\lambda + i\eta)$. With the same $\eta$ on both sides, any residual disagreement in the bulk is a property of the theory rather than of the smoothing.

\begin{figure*}[t]
\includegraphics[width=\textwidth]{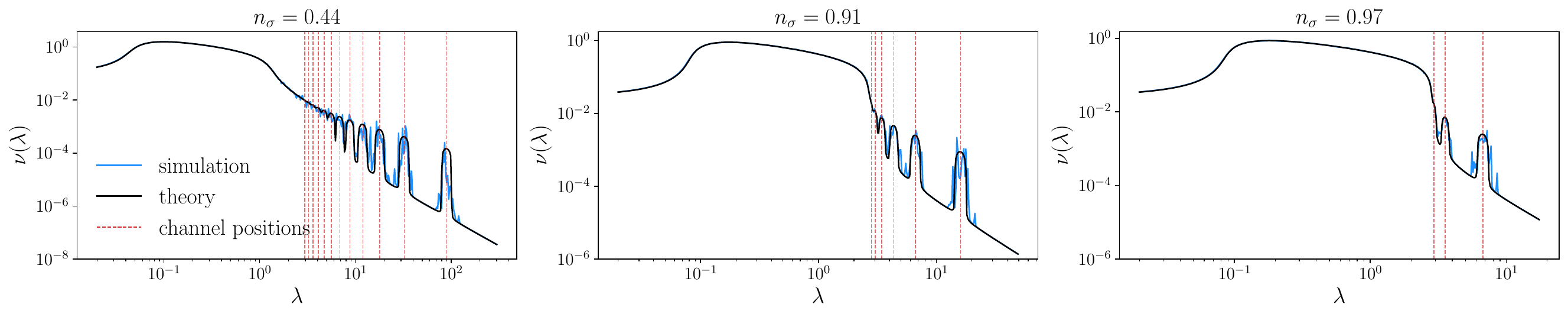}
\caption{Theory versus simulated densities across noise levels. Panels: noise levels $\sigma^2 \in \{2, 25, 75\}$ ($n_\sigma \approx 0.44, 0.91, 0.97$; $\mathrm{SNR} \approx 1.25, 0.10, 0.033$) at $\alpha = 1.5$. The noiseless baseline is the $\alpha = 1.5$ panel of Fig.~\ref{fig:overlay-alpha}. Blue: Lorentzian-smoothed empirical density. Black: density from the self-consistent equations. Dashed vertical lines: channel positions, the roots of the outlier condition, Eq.~\eqref{eq:outlier-noisy}. All panels at $N = 400$, $\ell = 300$, and $T = 800$ ($q = 0.5$ and $q_V = 0.75$). Both densities are smoothed at width $\eta = 0.01$. The empirical density is averaged over 80 independent realisations.
}
\label{fig:overlay-noise}
\end{figure*}

For $\alpha > 0$ the simulated densities in Fig.~\ref{fig:overlay-alpha} show a bulk that ends at its top in a hierarchy of isolated outlier eigenvalues. The number of outliers increases with increasing exponent value. In the bulk the theory curves closely follow the simulation for all values of $\alpha$. At the outliers, however, the comparison cannot be made directly. An outlier is a single eigenvalue, its position scatters between realisations at the relative scale $N^{-1/2}$~\cite{paul_2007, bai_yao_2008}, and the associated peak in the averaged empirical density therefore has a statistical width. In other words, the outliers are not self-averaging. The self-consistent equations, on the other hand, determine the self-averaging part of the spectrum. For outliers, the theory curves show a bump around every outlier position. More precisely, each latent channel enters the channel sums of Eq.~\eqref{eq:closure} through a term with its own pole, and for a channel separated in weight from its neighbours the solution turns this pole into a narrow peak that carries the mass of a single eigenvalue, and is centred on the channel's position (marked by the dashed lines). Its width is of the same order $N^{-1/2}$ as the scatter of the eigenvalue, which is why the theory and simulated peaks resemble each other. However, neither the shape nor the width of the theory bumps is an appropriate description of the outliers. The positions of the outliers and the law of their scatter are derived in Sec.~\ref{sec:edge}. Furthermore, where the hierarchy of outliers merges into the bulk, neither the smooth density nor the picture of isolated outliers applies. We return to this crossover region in the Discussion.
Figure~\ref{fig:overlay-noise} shows the comparison for $\alpha=1.5$ at finite noise, $\mathrm{SNR} \approx 1.25$, $0.10$, and $0.033$; the noiseless baseline, $\mathrm{SNR} = \infty$, is the $\alpha = 1.5$ panel of Fig.~\ref{fig:overlay-alpha}. Again, we find good agreement between theory and simulation, validating the self-consistent equations in their noisy form Eq.~\eqref{eq:closure-noisy}. As the noise level rises across the panels, a noise bulk emerges at order one while the outlier head is pulled inward by the signal fraction and thinned by the noise.
In Appendix~\ref{app:rankplots} we show the same comparisons as in Figures~\ref{fig:overlay-alpha} and~\ref{fig:overlay-noise} but represented as ranked eigenvalue spectra.

In Appendix~\ref{app:qv} we extend the comparison across the latent aspect ratio, from $q_V = 0.05$ to $q_V = 40$, again confirming the theory for the bulk. These values probe the two extremes of the latent aspect ratio: at $q_V = 0.05$ the data carry few latent features, $\ell \ll N$, and at $q_V = 40$ the latent hierarchy is far larger than the number of observed features, $\ell \gg N$. The latter is the regime likely relevant in applications: a measurement records a finite number of observed features, often a subset of a much larger set, while the system's underlying hierarchy of latent features can still be larger.

\subsection{Variational principle}
\label{sec:2PI}

The self-consistent equations in Eq.~\eqref{eq:closure-noisy} can be obtained from a variational principle
whose structure places the equations in the standard variational framework of many-body theory~\cite{luttinger_ward_1960, baym_kadanoff_1961}. The derivation applies to the latent sector the same two steps that Sec.~\ref{sec:exact} applied to the observation sector: a determinant linearised by a Gaussian vector, then reduced to its radial coordinate.
Writing the determinant in Eq.~\eqref{eq:F-factorised}, at general coupling $x$, instead of $a(g)$ which depends on the outer integration variable $g$, as a Gaussian integral over an auxiliary vector $\psi' \in \mathbb{R}^T$ linearises the dependence on $\mathbf{Y}$. Each column of $\mathbf{U}$ enters the exponent through a quadratic form that the Gaussian average computes exactly. The average over the columns of $\mathbf{U}$ produces the factor $\prod_m(1 - x\,w_m\,|\psi'|^2/T)^{-1/2}$, which depends on $\psi'$ only through its norm, and the radial variable $\phi := |\psi'|^2/T$ (below we will see that $\phi$ becomes the subordination scalar of Eq.~\eqref{eq:phi-def} at the saddle point) gives, with the convergence and continuation conventions of Sec.~\ref{sec:exact},
\begin{align}
\label{eq:inner-radial}
\mathbb{E}_U\Bigl[\det\bigl(\mathbf{I}_T - x\,\mathbf{Y}\bigr)^{-\frac12}\Bigr] &= c_T\int_0^\infty d\phi\;\phi^{T/2-1}\,e^{-T\phi/2}\nonumber\\
&\quad\times\prod_{m=1}^{\ell}\bigl(1 - x\,\phi\,w_m\bigr)^{-\frac12}\,,
\end{align}
where $c_T$ is a constant.
Substituting Eq.~\eqref{eq:inner-radial} at coupling $x = a$ into Eq.~\eqref{eq:Z-radial}, carrying the noise prefactor $(1 - b)^{-T/2}$ of Eq.~\eqref{eq:F-factorised} into the exponent, and collecting the exponents with $T/N = 1/q$ turns the generating function into a double radial integral,
\begin{align}
Z(z) &\propto \int_0^\infty\!dg\int_0^\infty\!d\phi\;\exp\Bigl(-\frac{N}{2}\,\mathcal{F}[g, \phi; z] + O(1)\Bigr)\,,
\end{align}
up to a $z$-independent constant in the exponent, with the two-variable free energy
\begin{align}
\label{eq:F-functional}
    \mathcal{F}[g, \phi; z] &= z g - \log g - \Phi[g, \phi]
\end{align}
and the interaction functional
\begin{align}
\begin{aligned}
\Phi[g, \phi] &= \frac{1}{q}\Bigl[\log\phi - (\phi - 1) - \log\bigl(1 - b(g)\bigr)\Bigr]\\
&\quad - \frac{q_V}{\ell}\sum_{m=1}^{\ell}\log\bigl(1 - a(g)\,\phi\,w_m\bigr)\,.
\end{aligned}
\end{align}
In many-body theory language $1/z$ is the bare propagator and $g$ the dressed propagator with all interaction effects resummed into it.
The structure of the free energy $\mathcal{F}[g, \phi; z]$ of Eq.~\eqref{eq:F-functional} is the Baym--Kadanoff form of many-body theory~\cite{luttinger_ward_1960, baym_kadanoff_1961} and includes: a bare coupling $zg$ between the spectral parameter and the propagator, a propagator entropy $-\log g$, and an interaction functional $\Phi$ that depends on the dressed propagators alone, such that $z$ enters nowhere else.
In many-body theory $\Phi$ is the Luttinger--Ward functional.

Joint stationarity of $\mathcal{F}$ reproduces the self-consistent equations. Stationarity in $g$, with self-energy $\Sigma = \partial\Phi/\partial g$, yields the Dyson equation,
and stationarity in $\phi$, $\partial\Phi/\partial\phi = 0$,
yields the $\phi$-equation, the first and second line of Eq.~\eqref{eq:closure-noisy}, respectively.
This confirms that the saddle value of the latent radial variable $\phi$ of Eq.~\eqref{eq:inner-radial} is the subordination scalar of Eq.~\eqref{eq:phi-def} at coupling $a$. The one-variable free energy of Eq.~\eqref{eq:free-energy} is this functional on the $\phi$-saddle, $\mathcal{F}(g) = \mathcal{F}[g, \phi_*(g); z]$ up to a constant, with $\phi_*(g)$ the solution of $\phi$ stationarity at fixed $g$.

\section{The weight continuum and the population baseline}
\label{sec:continuum}
In this section we prepare the analysis of the spectral shape in two steps. First, we take the continuum limit in the latent index $m$. The discrete channel sums of Eqs.~\eqref{eq:closure-noisy} and~\eqref{eq:closure} become integrals against the limiting measure of the channel weights, and regimes of qualitatively different spectral behaviours can be read off from that measure.
Second, we take the population limit $q \to 0$ as the baseline in which the mapping from weights to eigenvalues is most transparent. Both steps are taken at $\sigma = 0$ since noise enters the equations only through the scalar dressings $a$ and $b$, with no impact on the weight measure, and at $q = 0$ it only rescales and rigidly shifts the spectrum, cf. Eq.~\eqref{eq:q0-shift} and the surrounding discussion. 

\subsection{Continuum limit in the latent index}
\label{sec:continuum_latent}
The empirical measure $H_{\alpha,\ell}$ of the latent weights $\{w_m\}$ converges weakly (the fraction of weights in any fixed interval converges) as $\ell \to \infty$, and the character of the limit depends on the convergence of the latent-variance sum $S_\alpha$. Throughout, sub-critical, critical, and super-critical refer to the position of $\alpha$ relative to the convergence threshold of $S_\alpha$ at $\alpha = 1$. No connection to critical phenomena in physics is implied.

Sub-critically the limit is a non-degenerate measure $H_\alpha$, while super-critically the mass concentrates at zero and the leading weights escape to infinity. The large-$\ell$ regimes of $S_\alpha$ are collected in Appendix~\ref{app:salpha}. For sub-critical $\alpha < 1$, $S_\alpha \sim \ell^{1-\alpha}/(1 - \alpha)$ diverges and the weights form a continuum. Defining $x := m/\ell \in (0, 1]$ as the fractional latent feature rank, the weight profile in Eq.~\eqref{eq:w-def} becomes $w(x) = (1 - \alpha)\,x^{-\alpha}$. The fraction of channels with weight above $w$ is $x(w) = [(1 - \alpha)/w]^{1/\alpha} = H_\alpha((w, \infty))$, and differentiating in $w$ (with a minus sign, since $x(w)$ is the fraction above $w$) gives the normalised Pareto measure
\begin{align}
dH_\alpha(w) = \frac{(1 - \alpha)^{1/\alpha}}{\alpha}\,w^{-1 - 1/\alpha}\,dw\,,\quad w \ge 1 - \alpha\,,
\end{align}
with unit mean, and lower edge $1 - \alpha$. The measure has finite variance only below $\alpha = 1/2$, $\int dH_\alpha(w)\,w^2 = (1 - \alpha)^2/(1 - 2\alpha)$, so the sub-critical regime is split: for $\alpha < 1/2$ the weight measure is light-tailed and $\overline{w^2}$, the channel average of $w_m^2$, converges to the second moment $(1 - \alpha)^2/(1 - 2\alpha)$ of the measure, while for $1/2 < \alpha < 1$ it is heavy-tailed and $\overline{w^2}$ grows as $\ell^{2\alpha - 1}$, dominated by the leading channels. For critical $\alpha = 1$, $S_\alpha \sim \log\ell$ and the lower edge drifts to zero as $1/\log\ell$, so the critical point is the boundary case of the sub-critical construction. In this parametrisation the critical point $\alpha=1$ corresponds to Zipf's law, where latent variances decay as $m^{-1}$.

For super-critical $\alpha > 1$, $S_\alpha \to \zeta(\alpha)$ converges, and the leading channels carry weights $w_m \simeq \ell\,m^{-\alpha}/\zeta(\alpha)$ where $\zeta$ denotes the Riemann zeta function. The weights diverge with $\ell$, and so do their separations, since consecutive weights differ by $w_m - w_{m+1} = \ell\,\bigl[m^{-\alpha} - (m+1)^{-\alpha}\bigr]/\zeta(\alpha)$. At fixed $\ell$ each leading channel therefore retains a finite fraction $m^{-\alpha}/\zeta(\alpha)$ of the total variance and is separated from its neighbours by gaps of the order of the weights themselves. The top of the weight hierarchy consists of effectively finite-rank channels, while the ``deep'' channels form a continuum. In this regime the correct reading of the self-consistent equations at finite $\ell$ is a split. The top channels are treated as discrete outliers (see Sec.~\ref{sec:edge}) and the remainder as a bulk with a crossover region between the two.

In the continuum, the channel averages in Eqs.~\eqref{eq:closure} and~\eqref{eq:closure-noisy} become self-consistent integral equations,
\begin{align}
z - \frac{1}{g} &= \int_{1-\alpha}^{\infty}dH_\alpha(w)\,\frac{\phi\,w}{1 - (g/q_V)\,\phi\,w}\,,\nonumber\\
\frac{\phi - 1}{\phi} &= q\,g\int_{1-\alpha}^{\infty}dH_\alpha(w)\,\frac{w}{1 - (g/q_V)\,\phi\,w}\,,
\end{align}
of Mar\v cenko--Pastur type against $H_\alpha$~\cite{marchenko_pastur_1967, silverstein_bai_1995, silverstein_1995}, with the subordination scalar $\phi$.

\subsection{Population-limit baseline}
\label{sec:q0-baseline}

The baseline for the power-law structure of the spectrum is the population limit, $q \to 0$ ($T \to \infty$ at fixed $\ell$ and $N$), where the sampling noise disappears and the empirical covariance converges to its population counterpart.
For $q = 0$ the second line of the noiseless self-consistent equations in Eq.~\eqref{eq:closure} implies $\phi = 1$, and the Dyson equation reduces to
\begin{align}
\label{eq:q0-dyson}
z - \frac{1}{g} = \frac{1}{\ell}\sum_{m=1}^{\ell}\frac{w_m}{1 - (g/q_V)\,w_m}\,.
\end{align}
It is the self-consistent equation of $\mathbf{W} \to \mathbf{V}^T \mathbf{V}/\gamma^2$ (a coloured Wishart matrix at latent aspect ratio $q_V$) and was also derived in~\cite{paquette_phases_2024,fleig_2026}.
Notably, the population limit does not make the spectrum deterministic. The matrix $\mathbf{V}^T\mathbf{V}$ is a sum of $\ell$ random rank-one matrices, the outer products of the rows of $\mathbf{V}$, and the spectrum is spread around the rescaled weights $w_m/q_V$ over a range of order one. The map from weight to eigenvalue can be read off Eq.~\eqref{eq:q0-dyson} directly. The term of channel $m$ has a pole at $g = q_V/w_m$, where the free term is $1/g = w_m/q_V$ and the remaining channels contribute $\ell^{-1}\sum_{k \neq m} w_k\,(1 - w_k/w_m)^{-1} = 1 + O(\overline{w^2}/w_m)$. A weight that is well separated, $w_m \gg \max(q_V, \overline{w^2})$, therefore produces an isolated eigenvalue clear of the order-one bulk, at
\begin{align}
\label{eq:q0-spike-shift}
\lambda(w_m) = \frac{w_m}{q_V} + 1 + O(\overline{w^2}/w_m)
\end{align}
(at $q_V = 1$ this matches the classical spiked-Wishart position~\cite{baik_silverstein_2006}). All $\ell$ eigenvalues are identified with individual weights, $\lambda_m = w_m/q_V$, in the additional limit $q_V \to 0$ ($N \gg \ell$), where each latent direction is isolated and the order-one corrections are negligible. This is the limit of few latent features. In this case the top population eigenvalue sits at $w_1/q_V = N/S_\alpha$, which grows with the dimensions for every $\alpha > 0$, proportionally to $N$ for $\alpha > 1$ and to $N^{\alpha}$ for $\alpha < 1$. In contrast, the classical spiked models place spikes of fixed strength against the bulk~\cite{baik_silverstein_2006}.

Two structural features of the population limit control where the power law appears in the spectrum and where it fails in the sense that the ranked spectrum no longer follows the weight power law. 
First, the spectrum near zero is set by the aspect ratio $q_V$. The square case, $q_V=1$, divides the rank-deficient regime $q_V < 1$ from the gapped regime $q_V > 1$.
At $q_V < 1$ the spectrum carries a Dirac delta at zero (an atom of the spectral measure). The noiseless data matrix $\mathbf{U}\mathbf{V}$ has rank at most $\ell$, so the spectrum of $\mathbf{W}$ carries the delta with weight $1 - q_V$, and the Stieltjes transform $g$ carries the corresponding pole, $z\,g(z) \to 1 - q_V$ as $z \to 0$ below the nonzero support. The self-consistent equations contain the delta automatically, and in the limit of few latent features $q_V \to 0$ the spectrum degenerates to this delta plus $\ell$ isolated eigenvalues at the positions $w_m/q_V$.

Second, the power law survives only at the top. By Eq.~\eqref{eq:q0-spike-shift}, the ranked spectrum tracks $w_{m}/q_V$, up to an additive $O(1)$ shift, where $w_{m}/q_V \gg 1$, and departs from the power law where the weights descend into the order-one bulk. Through the same weight-to-eigenvalue map, the density inherits the Pareto tail of the weight measure, $\nu_0(\lambda) \sim \lambda^{-1 - 1/\alpha}$ at large $\lambda$, whose variance is finite exactly in the light-tailed regime $\alpha < 1/2$.
The breakdown of the power law at the bottom is thus an intrinsic feature of the aspect ratio $q_V$, present already at $q = 0$. Finite $q$ modifies it quantitatively through $\phi \ne 1$ but does not create it.

\section{The top of the spectrum: outliers}
\label{sec:edge}

\begin{figure*}[t]
\includegraphics[width=0.49\textwidth]{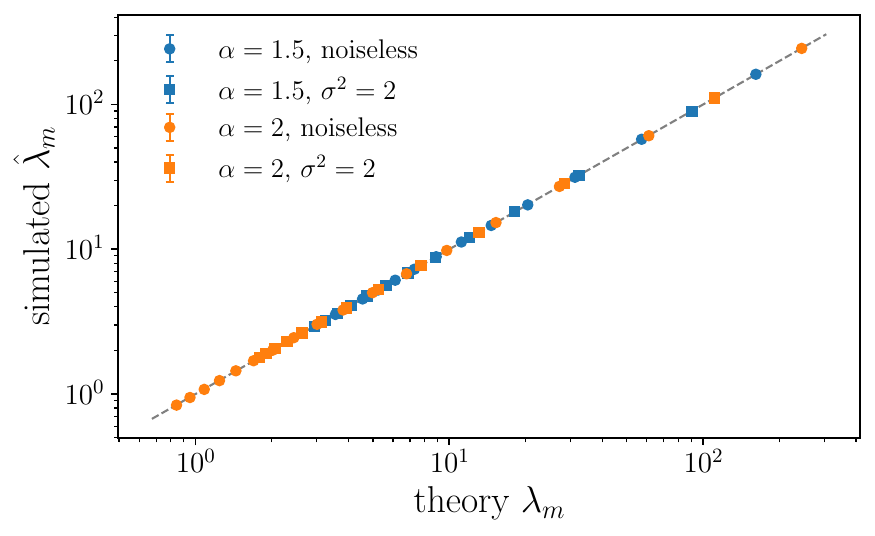}\hfill
\includegraphics[width=0.49\textwidth]{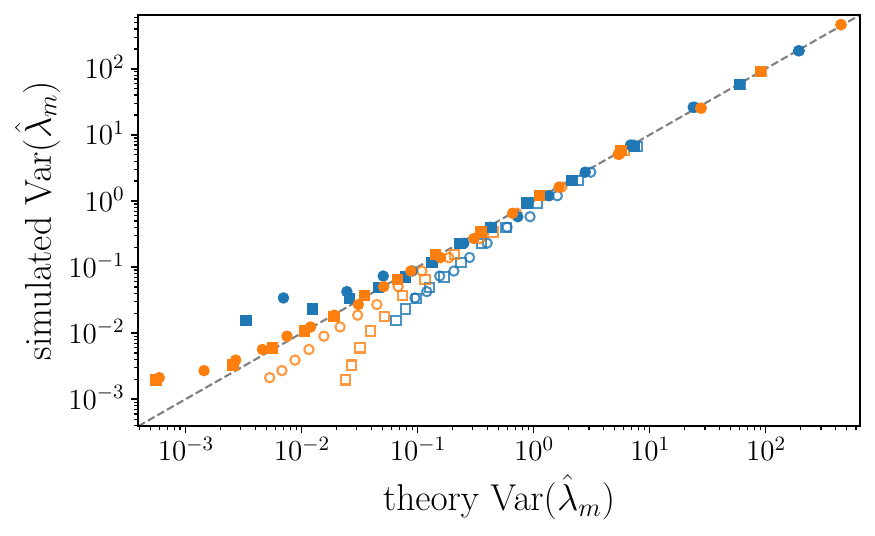}
\caption{Theory versus simulation at the top of the spectrum. Positions (left) and fluctuations (right) of the outlier eigenvalues, the dashed lines marking equality. Left: theory position from the outlier condition, Eq.~\eqref{eq:outlier-noisy}, versus the simulated rank-wise mean, for all channel-tracked ranks of four ensembles: $\alpha \in \{1.5, 2\}$ (blue and orange, respectively), noiseless (circles) and noisy at $\sigma^2 = 2$ (squares). Right: theory versus simulated variance of the top eigenvalues, for the same ensembles and ranks in the same conventions. Filled markers: the fluctuation law, Eq.~\eqref{eq:outlier-fluct}. Open markers: its far-outlier limit, Eq.~\eqref{eq:far-outlier-var}. All ensembles at $N = 400$, $\ell = 300$, and $T = 800$ ($q = 0.5$ and $q_V = 0.75$), with means and variances taken over $R = 400$ realisations. Error bars mark the Monte Carlo standard error of the mean (left; smaller than the markers throughout) and the variance-estimator standard error $\sqrt{2/R}$ times the prediction (right).}
\label{fig:outlier-check}
\end{figure*}

The top of the spectrum carries structure beyond a smooth density. The leading channels eject isolated outlier eigenvalues. Each is a single eigenvalue that sits at a deterministic position and fluctuates around it at the relative scale $N^{-1/2}$. As Sec.~\ref{sec:checks} showed, the leading-order equations stay finite at an outlier, placing there a narrow peak with the mass and the position of the eigenvalue. Inside that peak, however, the spectrum is a single fluctuating eigenvalue, which is not self-averaging, and its law is not determined by the equations alone. In this section we compute the outlier positions from the leading-order equations and derive the fluctuation law around them.

\subsection{Outliers of the leading channels}
\label{sec:outliers}

A latent channel whose weight is sufficiently separated from its neighbours acts as a finite-rank perturbation that ejects an eigenvalue from the bulk formed by the other channels. The separation is strongest at super-critical $\alpha > 1$, where the top weights $w_m \simeq \ell m^{-\alpha}/\zeta(\alpha)$ are separated by gaps of the order of the weights themselves, cf. Sec.~\ref{sec:continuum_latent}. The ratios of consecutive weights are however fixed at every $\alpha > 0$, $w_m/w_{m+1} = \bigl((m+1)/m\bigr)^{\alpha}$, independent of $\ell$, so the leading weights retain a finite relative separation at any finite $\ell$. Sub-critically this separation does not survive the continuum limit, in which the weight measure of Sec.~\ref{sec:continuum_latent} has no gaps. In this sense the sub-critical separation is a finite-size property.

The outliers are identified by the resonance of the per-channel quadratic form Eq.~\eqref{eq:channel-qform}, probed at the dressed coupling $a$ of Eq.~\eqref{eq:ab-def}.
The quadratic form diverges when $a(\lambda)\,\tilde\phi(\lambda)\,w_m = 1$, that is,
\begin{align}
\label{eq:outlier-noisy}
s\,\tilde g(\lambda)\,\tilde\phi(\lambda)\,w_m = q_V\,\bigl(1 - q\,n_\sigma\,\tilde g(\lambda)\bigr)\,,
\end{align}
where $(\tilde g, \tilde\phi)$ is the real bulk solution of Eq.~\eqref{eq:closure-noisy} with the term of channel $m$ left out of the channel sums. The channel removal is required by what the outlier condition, Eq.~\eqref{eq:outlier-noisy}, demands: an isolated eigenvalue is a pole of the resolvent of $\mathbf{W}$, and a pole can only be located through an object that stays regular at it, i.e. the resolvent of the bulk formed by the other channels. With channel $m$ kept in the sums, the equations develop the channel's own resonance peak around $\lambda_m$ and the sharp root disappears.

We define an outlier as a point for which the condition given by Eq.~\eqref{eq:outlier-noisy} has an isolated root. Isolated means that the root lies in a gap of the channel-removed bulk, whose density vanishes in a neighbourhood of the root. We refer to such eigenvalues as \emph{channel-tracked}. The isolated roots can be located numerically by the root finder described in Appendix~\ref{app:solver}. At finite $\ell$ the condition has isolated roots at every $\alpha > 0$, and the spectrum ends in a channel-tracked hierarchy. At the simulated sizes the count grows with the exponent, $0, 1, 5, 6, 8, 13, 17$ at $\alpha = 0, 0.3, 0.6, 0.75, 1, 1.5, 2$, respectively, in the noiseless case. In Fig.~\ref{fig:overlay-alpha} the outlier positions are marked by the dashed vertical lines. The regimes differ in how the hierarchy scales. At super-critical $\alpha$ the separations of the leading eigenvalues grow with the dimensions, while at sub-critical $\alpha$ the top ranks remain marginally separated at every size.
We note that the classical Baik--Ben Arous--P\'ech\'e (BBP) threshold for a single spike~\cite{baik_benarous_peche_2005, baik_silverstein_2006} does not apply to our model.
We return to this point in the Discussion.

Since noise lowers the resonance product at fixed $\lambda$ through the prefactor $s < 1$, while the opposing screening correction is negligible above the bulk, the number of outliers falls with the noise level. In Fig.~\ref{fig:overlay-noise}, the outlier positions are marked by the dashed vertical lines. We note that Eq.~\eqref{eq:outlier-noisy} has the form of the rigorous low-rank perturbation theory for rectangular matrices~\cite{benaych_nadakuditi_2012}, and the outlier theory of additive signal-plus-noise matrices at extensive rank is developed in~\cite{landau_mel_ganguli_2023}.

Far above the bulk, $\tilde g \to 1/\lambda$ and $\tilde\phi \to 1$, and inserting these limits into Eq.~\eqref{eq:outlier-noisy} gives $s\,w_m/\lambda = q_V(1 - q\,n_\sigma/\lambda)$, hence $\lambda_m = s\,w_m/q_V + q\,n_\sigma + O(1)$, where the remaining $O(1)$ correction collects the bulk interaction. Thus, a far outlier sits at its weight, rescaled by the signal fraction and the latent aspect ratio, and shifted by the noise background.
In terms of the defining parameters of the model the rescalings cancel: $s\,w_m/q_V = N\,m^{-\alpha}/\gamma^2$, so the far outlier sits at $\lambda_m = N\,m^{-\alpha}/\gamma^2 + q\,n_\sigma + O(1)$, which is $N$ times the channel's per-coordinate variance, shifted by the noise background. The latent dimension enters only through the normalisation $\gamma^2$. At super-critical $\alpha$, where $S_\alpha$ converges, the discrete top at fixed $N$ is therefore insensitive to $q_V$, which shapes the spectrum through the bulk instead.

Figure~\ref{fig:outlier-check} (left) compares the theory positions, the roots of Eq.~\eqref{eq:outlier-noisy}, with the simulated rank-wise means, for every channel-tracked rank of four ensembles, $\alpha \in \{1.5, 2\}$, noiseless and at $\sigma^2 = 2$, with $R = 400$ realisations each. All points fall on the diagonal within a few Monte Carlo standard errors of the means. The comparison protocol is recorded in Appendix~\ref{app:numerics}.

\subsection{Outlier fluctuations}
\label{sec:outlier-fluct}

The outlier positions derived so far are deterministic. Next, we derive the law of the fluctuations around them, a Gaussian at relative scale $N^{-1/2}$ as in the fixed-spike setting~\cite{bai_yao_2008}. We state and explain the law in closed form for our model. Details of the somewhat lengthy derivation are given in Appendix~\ref{app:fluct}.

The fluctuation law is derived by treating channel $m$ as a perturbation of the bulk formed by the other channels, and every quantity entering it is therefore evaluated on that channel-removed bulk. We write $\tilde g$ for the bulk solution of the self-consistent equations Eq.~\eqref{eq:closure-noisy} without channel $m$ (the removal of Sec.~\ref{sec:outliers}), evaluated at $\lambda_m$. Together with $\tilde \varphi := 1 + q(\lambda_m \tilde g - 1)$, which equals $\tilde\phi/(1 - q\,n_\sigma\,\tilde g)$ by Eq.~\eqref{eq:phi-identity-noisy} and at $\sigma = 0$ coincides with the subordination scalar $\tilde\phi$,
the law reads
\begin{align}
\label{eq:outlier-fluct}
\begin{aligned}
\hat\lambda_m &= \lambda_m + \sqrt{\frac{v_m}{N}}\;\frac{\xi_m}{\bigl|\partial_\lambda(\tilde g\tilde \varphi)\bigr|_{\lambda_m}}\,,\quad \xi_m \sim \mathcal{N}(0, 1)\,,\\
v_m &= 2\,\tilde \varphi^2\,\tau_2 + 2\,q^2 \tilde g^2\bigl(\lambda_m^2\tau_2 - 2\lambda_m \tilde g + 1\bigr)\\
&\quad + 4\,q\,\tilde g\tilde \varphi\bigl(\lambda_m\tau_2 - \tilde g\bigr) + 2\,q\,\tilde g^2\bigl(2\tilde \varphi - 1\bigr)\,,
\end{aligned}
\end{align}
where $\tau_2 := -\tilde g'(\lambda_m)$ is the average of $(\lambda_m - x)^{-2}$ over the bulk spectral density, and $\hat\lambda_m$ is the realised outlier eigenvalue.
Every quantity in Eq.~\eqref{eq:outlier-fluct} is an output of the self-consistent equations, so the fluctuation law is available at every outlier whose position Eq.~\eqref{eq:outlier-noisy} predicts. Together with the position, the law completes the channel-tracked description of the discrete top of the spectrum. In Appendix~\ref{app:fluct} we derive the far-outlier limit of the law, which connects it to the classical spiked-covariance fluctuations and serves as a check supporting it.

Figure~\ref{fig:outlier-check} (right) compares the law with the sample variances of the top eigenvalues, over the same ensembles and ranks as the position comparison. The law holds over most of the hierarchy, within a few times the relative variance-estimator standard error $\sqrt{2/R}$. For the far outliers the law merges with its far-outlier limit, Eq.~\eqref{eq:far-outlier-var}, and the paired filled and open markers merge. Further down the full law remains accurate while the far-outlier limit visibly departs. At the lowest channel-tracked ranks, closest to the bulk, the simulation departs from the law itself, and the measured variance exceeds the prediction. The departure is explained by an approximation in the derivation. The law is derived at fixed bulk: the fluctuations of the channel's own directions are propagated through the characteristic equation of Appendix~\ref{app:fluct}, while the bulk formed by the other channels enters only through its deterministic resolvent. Closest to the bulk this approximation fails first. The fluctuations of the neighbouring bulk eigenvalues are no longer negligible against the gap that isolates the root, and their neglected contribution adds variance. The same comparison at sub-critical $\alpha$ is recorded in Appendix~\ref{app:numerics}.

\section{Discussion}
\label{sec:discussion}

We have introduced the power-law latent feature model with noise and provided essential first steps of a spectral theory for the model, with self-consistent equations for the eigenvalue density, and an analysis of outliers, both validated against simulations.

In this work we focused on theory development and leave a comparison of the model with data for future work. While we believe that the model captures key characteristics of power-law spectra, we expect most real data to carry additional structure not captured by our model.
In particular, the model makes two main idealisations. The first is to assume a perfect power law. Yet, empirical spectra are often not a perfect power law, and their local slope may drift with rank, so that no single exponent $\alpha$ holds across all scales.
The second is the isotropy of noise which adds the same variance $\sigma^2$ to every observed direction.
Structured noise, e.g. heteroscedastic, low-rank, or spatially correlated noise, lifts different directions of observation space by different amounts.

The theory we have developed in this work is model specific: the outlier positions and their fluctuation law are derived from the model's own equations rather than from classical outlier laws~\cite{baik_benarous_peche_2005, baik_silverstein_2006, paul_2007, bai_yao_2008}.
The reason is that such theorems assume either a population spectrum bounded uniformly in the dimension or finitely many spikes of fixed strength, which our model violates, since its leading population eigenvalues grow with the dimension for every $\alpha > 0$ (Sec.~\ref{sec:q0-baseline}). The bulk theory is unaffected, however. Sub-critically the deterministic equivalents require only weak convergence of the weight measure~\cite{silverstein_bai_1995, silverstein_1995}, and super-critically the equations are read at finite $\ell$, with the top channels split off as outliers (Sec.~\ref{sec:continuum_latent}).

There are several directions for future work. First, the theory developed in this work provides information on the required number of samples and features in a measurement, at a given noise level, for a hypothesised power law to be resolvable. Such a quantitative theory for the resolution of power laws in empirical covariances may guide the statistics of data collection in experiments.

Second, the description of the crossover region between the bulk and the hierarchy of outliers is left open. Classical analyses of this transition concern the critical regime of the BBP transition~\cite{baik_benarous_peche_2005}, in which a spike of near-critical strength stands against a bulk with a fixed edge. In our model the crossover is of a different kind. Nothing in the comparison is fixed: a channel entering the crossover is measured against a bulk whose top is built from its neighbours in the hierarchy, with weights of the same order as its own. The crossover is therefore a collective problem of many channels of comparable weight. Its signature is visible in Sec.~\ref{sec:outlier-fluct}: the variance of the lowest channel-tracked eigenvalues exceeds the fluctuation law, which holds the bulk at its deterministic resolvent. A quantitative theory of this crossover region is beyond the scope of this work, and we leave it as an open problem.

Finally, we note that the theory affords a diagrammatic interpretation.
The self-consistent equations admit standard planar-diagram derivations~\cite{thooft_1974, burda_signal_noise_2004, pennington_worah_2017, baym_1962}, and the functional $\Phi$ is a two-particle irreducible functional~\cite{cornwall_jackiw_tomboulis_1974,benedetti_gurau_2018}.
Furthermore, the noise fits a similar picture with noise contractions organising into rainbow diagrams whose resummation produces the screening structure of the noisy self-energy of Eq.~\eqref{eq:closure-noisy}~\cite{brezin_zee_1994, burda_signal_noise_2004}.

\section*{Acknowledgements}
The author is grateful to Ilya Nemenman for discussions in which the idea for the power-law latent feature model emerged.

\appendix

\section{Regimes of the partial sum $S_\alpha$}
\label{app:salpha}

The partial sum $S_\alpha = \sum_{m=1}^{\ell} m^{-\alpha}$ is a generalised harmonic sum, and its large-$\ell$ behaviour is classical~\cite{apostol_1976}, with three regimes,
\begin{align}
S_\alpha &= \frac{\ell^{1-\alpha}}{1-\alpha} + \zeta(\alpha) + O(\ell^{-\alpha})\,,\quad 0 \le \alpha < 1\,,\nonumber\\
S_1 &= \log\ell + \gamma_E + O(\ell^{-1})\,,\quad \alpha = 1\,,\nonumber\\
S_\alpha &= \zeta(\alpha) + O(\ell^{1-\alpha})\,,\quad \alpha > 1\,,
\end{align}
where $\gamma_E = 0.57721\ldots$ is the Euler--Mascheroni constant and $\zeta$ the Riemann zeta function, which enters the first line through its analytic continuation to $\alpha < 1$~\cite{nist_dlmf}. These regimes underlie the sub-critical, critical (Zipf point $\alpha=1$), and super-critical spectral regimes of this work. The moments of the channel weights follow as $\overline{w^k} = \ell^{k-1} S_{k\alpha}/S_\alpha^k$, so the regimes above, applied at exponent $k\alpha$, place the moment thresholds at $\alpha = 1/k$. In particular, $\overline{w^2}$ converges for $\alpha < 1/2$ and grows as $\ell^{2\alpha - 1}$ for $1/2 < \alpha < 1$.

\section{Deterministic equivalents of the coloured Wishart}
\label{app:silverstein}
We derive Eqs.~\eqref{eq:phi-eq} and~\eqref{eq:K-eq} at general coupling $x$. The result is standard~\cite{silverstein_bai_1995, silverstein_1995}. All expectations in this appendix are over $\mathbf{U}$. Let $u_m \in \mathbb{R}^T$ be the $m$-th column of $\mathbf{U}$, so the columns are i.i.d., $u_m \sim \mathcal{N}(0, \mathbf{I}_T)$, and $\mathbf{Y} = T^{-1}\sum_m w_m u_m u_m^T$. We define its resolvent $\mathbf{R}_Y := (\mathbf{I}_T - x \mathbf{Y})^{-1}$. The resolvent identity gives
\begin{align}
\mathbf{R}_Y = \mathbf{I}_T + \frac{x}{T}\sum_{m=1}^{\ell} w_m\,u_m u_m^T\,\mathbf{R}_Y\,.
\end{align}
We first average the identity. We use the identity $\mathbb{E}[(u_m)_a F(\mathbf{U})] = \mathbb{E}[\partial F/\partial(u_m)_a]$ for Gaussian integration-by-parts, and apply it to the entry $(a, b)$ of $\mathbb{E}[u_m u_m^T\mathbf{R}_Y]$ (with $a$, $b=1,\ldots, T$), i.e. to $F = (u_m^T\mathbf{R}_Y)_b$. This yields
\begin{align}\label{eq:uuR}
\mathbb{E}\bigl[u_m u_m^T\mathbf{R}_Y\bigr] &= \mathbb{E}[\mathbf{R}_Y] + \frac{x\,w_m}{T}\,\mathbb{E}\bigl[(u_m^T\mathbf{R}_Y\,u_m)\,\mathbf{R}_Y\bigr]\nonumber\\
&\quad + \frac{x\,w_m}{T}\,\mathbb{E}\bigl[\mathbf{R}_Y\,u_m u_m^T{\mathbf{R}_Y}^{\,T}\bigr]\,,
\end{align}
where $\mathbb{E}[\mathbf{R}_Y]$ comes from the derivative acting on the explicit factor $u_m^T$, and the other two terms from its implicit action on the resolvent, through
\begin{align}
\frac{\partial \mathbf{R}_Y}{\partial (u_m)_a} = \frac{x\,w_m}{T}\,\mathbf{R}_Y\bigl(e_a u_m^T + u_m e_a^T\bigr)\mathbf{R}_Y\,,
\end{align}
with $e_a$ the $a$-th standard basis vector of $\mathbb{R}^T$. Now we take the trace of both sides of Eq.~\eqref{eq:uuR}. The second term on the right-hand side carries two factors of order $T$, the quadratic form $u_m^T\mathbf{R}_Y\,u_m$ and the trace of $\mathbf{R}_Y$. The last term carries a single quadratic form only and is neglected, at relative order $1/T$. Furthermore, the trace self-averages, $\mathrm{tr}\,\mathbf{R}_Y = T\phi + O(1)$ with $\phi := T^{-1}\mathbb{E}[\mathrm{tr}\,\mathbf{R}_Y]$, so that the average of the product in the second term factorises at leading order. The traced identity thus becomes a linear equation,
\begin{align}
\mathbb{E}[u_m^T\mathbf{R}_Y\,u_m] = T\phi + x\,\phi\,w_m\,\mathbb{E}[u_m^T\mathbf{R}_Y\,u_m]\,,
\end{align}
whose solution is
\begin{align}
\label{eq:channel-qform}
\mathbb{E}\bigl[u_m^T\mathbf{R}_Y\,u_m\bigr] &= \frac{T\,\phi}{1 - x\,\phi\,w_m}\,.
\end{align}
This quadratic form is the resolvent probed along the direction of channel $m$, and its divergence at $x\,\phi\,w_m = 1$ signals an eigenvalue aligned with that channel. The outlier condition of Sec.~\ref{sec:outliers} is this resonance, evaluated on the channel-removed bulk solution. Equivalently, writing $\mathbf{Y} = \mathbf{Y}^{(m)} + (w_m/T)\,u_m u_m^T$ for the matrix without channel $m$, the Sherman--Morrison formula~\cite{potters_bouchaud_2020} gives $u_m^T\mathbf{R}_Y\,u_m = u_m^T\mathbf{R}^{(m)}\,u_m/[1 - (x w_m/T)\,u_m^T\mathbf{R}^{(m)}\,u_m]$, and since $u_m$ is independent of $\mathbf{R}^{(m)}$ this form holds uniformly in $w_m$. This justifies evaluating the resonance on the channel-removed bulk solution.
Substituting this solution into the untraced identity, Eq.~\eqref{eq:uuR}, gives
\begin{align}
\mathbb{E}\bigl[u_m u_m^T\mathbf{R}_Y\bigr] &= \frac{\mathbb{E}[\mathbf{R}_Y]}{1 - x\,\phi\,w_m}\,.
\end{align}
Taking the normalised trace of the resolvent identity and inserting Eq.~\eqref{eq:channel-qform} gives
\begin{align}
\phi = \Biggl(1 - \frac{x}{T}\sum_m\frac{w_m}{1 - x\,\phi\,w_m}\Biggr)^{-1}\,,
\end{align}
which is Eq.~\eqref{eq:phi-eq} on the branch $\phi \to 1$ as $x \to 0$. For $\mathcal{K}$, writing $\mathrm{tr}[\mathbf{Y}\mathbf{R}_Y] = T^{-1}\sum_m w_m\,u_m^T\mathbf{R}_Y\,u_m$ and inserting Eq.~\eqref{eq:channel-qform}, we obtain
\begin{align}
\mathcal{K}(x) = \frac{1}{\ell}\sum_m\frac{w_m}{T}\,\mathbb{E}\bigl[u_m^T\mathbf{R}_Y\,u_m\bigr]
= \frac{1}{\ell}\sum_{m=1}^{\ell}\frac{\phi\,w_m}{1 - x\,\phi\,w_m}\,,
\end{align}
which is Eq.~\eqref{eq:K-eq}.

\section{The closed-form branch at $\alpha = 0$}
\label{app:alpha0}

In this appendix we derive the exact $\alpha = 0$ reductions of the self-consistent equations, noiseless and noisy, and connect them to the free-probability results of~\cite{fleig_nemenman_2022}.

\subsection{Noiseless equation at $\alpha = 0$}
We derive the $\alpha = 0$ reduction of the self-consistent equations, Eq.~\eqref{eq:closure}.
At $\alpha = 0$ all latent weights are $w_m = 1$ and the channel sums in Eq.~\eqref{eq:closure} are trivial, reducing the self-consistent equations to a polynomial system
\begin{align}
z - \frac{1}{g} &= \frac{\phi}{1 - \frac{g\,\phi}{q_V}}\,,\nonumber\\
(\phi - 1)\Bigl(1 - \frac{g\,\phi}{q_V}\Bigr) &= q\,g\,\phi\,.
\label{eq:closure-alpha0}
\end{align}
The natural variable for eliminating $\phi$ is the product $u := zg$. The identity, Eq.~\eqref{eq:phi-identity-noisy}, at $n_\sigma = 0$ gives $\phi = 1 + q(u - 1)$, which is a function of $u$ alone, so the physical branch is rational in $u$. Eliminating $\phi$ through $u$ yields the cubic,
\begin{align}
\label{eq:cubic}
0 &= q\,z^2 g^3 + (q\,q_V + 1 - 2q)\,z\,g^2\nonumber\\
&\quad + \bigl[(q_V - 1)(1 - q) - q_V z\bigr]g + q_V\,.
\end{align}
This cubic matches the free-probability result of~\cite{fleig_nemenman_2022}: Eq.~(C24) of~\cite{fleig_nemenman_2022}, with the change of variables $g = q_V\,\mathfrak h + (1 - q_V)/z$, Eq.~(C15) of~\cite{fleig_nemenman_2022}, where $\mathfrak h$ is the Stieltjes transform of the nonzero spectrum alone, and the second term is the zero-eigenvalue delta of Sec.~\ref{sec:q0-baseline}.

\subsection{Noisy equation at $\alpha = 0$}
\label{app:alpha0-noisy}

At finite noise the self-consistent equations remain algebraic at $\alpha = 0$, and the resulting polynomial provides the exact cross check with the finite-SNR results of~\cite{fleig_nemenman_2022}. Setting $w_m \equiv 1$ in Eq.~\eqref{eq:closure-noisy} and eliminating $\phi$ through the noisy identity, Eq.~\eqref{eq:phi-identity-noisy}, leaves a single polynomial equation for the transform $g$, the quartic
\begin{align}
\label{eq:quartic-noisy}
0 &= n_\sigma s\,q^2 z^2\,g^4 - s\,q\,z\bigl[z - 2 n_\sigma(1 - q)\bigr]g^3\nonumber\\
&\quad + \bigl[n_\sigma s\,(1 - q)^2 - q\,q_V\,z - (1 - 2q)\,s\,z\bigr]g^2\nonumber\\
&\quad + \bigl[q_V\,(z - 1 + q) + (1 - q)\,s\bigr]g - q_V\,.
\end{align}
At $n_\sigma = 0$ the leading coefficient vanishes and the equation reduces to the noiseless cubic Eq.~\eqref{eq:cubic}.

The finite-SNR treatment of~\cite{fleig_nemenman_2022}, whose SNR coincides with Eq.~\eqref{eq:snr-def} at $\alpha = 0$, offers comparisons in each of its two limits. In the classical statistics (population) limit ($q \to 0$ at fixed $q_V$) the noisy spectrum of that work is the noiseless spectrum rescaled by the signal fraction $s$ and shifted by the noise fraction $n_\sigma$. Our equations obey the same relation: in the $q \to 0$ limit the quartic, Eq.~\eqref{eq:quartic-noisy}, reduces to the rescaled and shifted noiseless cubic, so the two spectra coincide exactly.
At the level of the limiting Stieltjes transform the comparison is the identity
\begin{align}
\label{eq:q0-shift}
g^{(\sigma)}(z)\big|_{q = 0} = \frac{1}{s}\,g^{(0)}\Bigl(\frac{z - n_\sigma}{s}\Bigr)\,,
\end{align}
where $g^{(0)}$ is the transform of the noiseless model ($s = 1$, $n_\sigma = 0$). The identity holds at every $\alpha$. By this identity, the exact spectrum at $q \to 0$ is the $s$-rescaled signal spectrum rigidly shifted by $n_\sigma$. The shift is the collapsed noise bulk that becomes the noise floor.
In the intensive limit ($q$ fixed and $q_U \to 0$) the construction of~\cite{fleig_nemenman_2022} adds the signal and noise sample-covariance blocks by free additive convolution.
This neglects the signal--noise cross terms and the $O(q_U)$ term of the signal factor, both of which vanish in that limit. At finite $q_U$ the difference is exactly this neglected content, which our equations keep.

\section{Second-moment test of the self-consistent equations}
\label{app:moments}
Moments of $\mathbf{W}$ can be computed by Wick contractions~\cite{potters_bouchaud_2020} without input from the self-consistent equations. The lowest informative moment is $m_2 := N^{-1}\mathbb{E}[\mathrm{tr}\,\mathbf{W}^2]$. Expanding the self-energy of the Dyson form of Eq.~\eqref{eq:closure-noisy-raw}, $\Sigma = z - 1/g$, in the moment expansion $g = z^{-1} + m_1 z^{-2} + m_2 z^{-3} + \ldots$ gives $\Sigma = m_1 + (m_2 - m_1^2)\,g + O(g^2)$, with $m_1 = 1$ by the trace normalisation. Exact computation by Wick contraction gives, at leading order in the dimensions and at finite $\ell$,
\begin{align}
\label{eq:m2}
m_2 = 1 + q + s^2\,\frac{\overline{w^2}}{q_V}\,.
\end{align}
On the other hand, expanding the self-consistent Eq.~\eqref{eq:closure-noisy} at small $g$ gives $\Sigma = 1 + g\,(q + s^2\,\overline{w^2}/q_V) + O(g^2)$, showing that the second moment is in agreement with Eq.~\eqref{eq:m2}. 

\begin{figure*}[t]
\centering
\begin{minipage}[t]{0.49\textwidth}
\centering
\includegraphics[width=\textwidth]{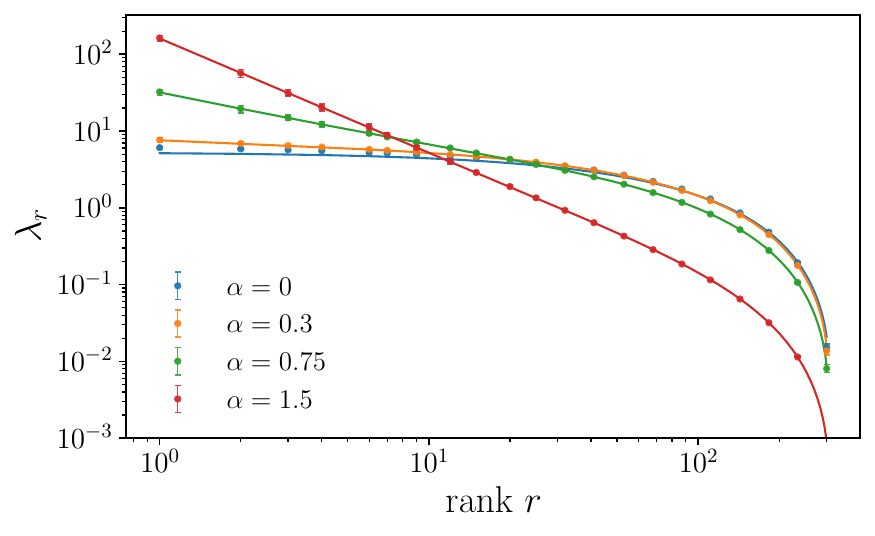}\\
(a) Across the exponent (noiseless)
\end{minipage}\hfill
\begin{minipage}[t]{0.49\textwidth}
\centering
\includegraphics[width=\textwidth]{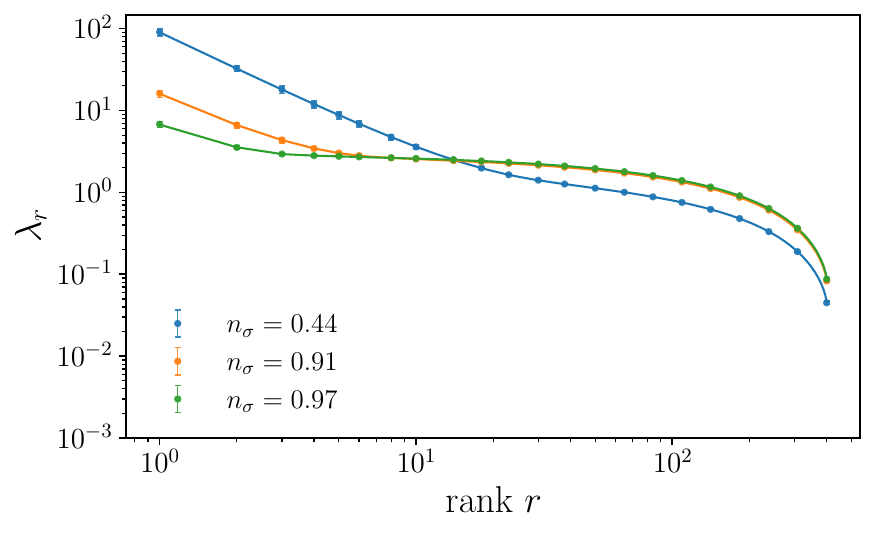}\\
(b) Across noise levels at $\alpha = 1.5$
\end{minipage}
\caption{Ranked-spectrum version of the density comparisons of Figs.~\ref{fig:overlay-alpha} and~\ref{fig:overlay-noise}, at the same parameters. (a) The noiseless ensembles of Fig.~\ref{fig:overlay-alpha}, $\alpha \in \{0, 0.3, 0.75, 1.5\}$. (b) The noise levels of Fig.~\ref{fig:overlay-noise}, $\sigma^2 \in \{2, 25, 75\}$ at $\alpha = 1.5$; the noiseless $\alpha = 1.5$ curve is in (a). Lines are the theory curves of Eq.~\eqref{eq:rank-curve}. Points are the simulated rank-wise means at log-spaced ranks, in the conventions of Fig.~\ref{fig:hook-ranks}. Means over 80 independent realisations. Error bars mark the 10th--90th percentile range over realisations.}
\label{fig:ranked-overlays}
\end{figure*}

\section{Numerical solver algorithm}
\label{app:solver}

The numerical solver computes the pair $(g, \phi)$ of Eq.~\eqref{eq:closure-noisy} at each complex spectral parameter $z = \lambda + i\eta$ with $\eta > 0$.
Algorithm~\ref{algo:solver} below defines the fixed-point solver.

\begin{algorithm}[H]
\caption{Damped joint fixed point for $(g, \phi)$}\label{algo:solver}
\KwIn{$z = \lambda + i\eta$ with $\eta > 0$; ratios $q$, $q_V$; weights $\{w_m\}_{m=1}^{\ell}$; fractions $s$, $n_\sigma$; tolerance $\varepsilon$; damping $\theta \in (0, 1]$; iteration cap $n_{\mathrm{max}}$; optional warm start $(g_{\mathrm{init}}, \phi_{\mathrm{init}})$}
\KwOut{the pair $(g, \phi)$ and the smoothed density $\nu_\eta(\lambda)$}
\KwInit{$q_U \leftarrow q\,q_V$; $g \leftarrow g_{\mathrm{init}}$ if provided, else $g \leftarrow 1/z$; if $\mathrm{Im}\,g > 0$ then $g \leftarrow \bar g$; $\phi \leftarrow \phi_{\mathrm{init}}$ if provided, else $\phi \leftarrow 1$}
\Repeat{$|\Delta g| < \varepsilon\,\max(|g|, 1)$ \emph{and} $|\Delta\phi| < \varepsilon\,\max(|\phi|, 1)$, \emph{or} the iteration count exceeds $n_{\mathrm{max}}$ (no convergence)}{
  $b \leftarrow q\,n_\sigma\,g$;\quad $a \leftarrow s\,g/\bigl(q_V(1 - b)\bigr)$\;
  $\phi_{\mathrm{new}} \leftarrow \Bigl[1 - q_U\,a\,\dfrac{1}{\ell}\sum_m \dfrac{w_m}{1 - a\,\phi\,w_m}\Bigr]^{-1}$\;
  $\Sigma \leftarrow \dfrac{n_\sigma}{1 - b} + \dfrac{s}{(1 - b)^2}\,\dfrac{1}{\ell}\sum_m \dfrac{\phi\,w_m}{1 - a\,\phi\,w_m}$\;
  $\Delta\phi \leftarrow \theta\,(\phi_{\mathrm{new}} - \phi)$;\quad $\phi \leftarrow \phi + \Delta\phi$\;
  $g_{\mathrm{new}} \leftarrow 1/(z - \Sigma)$\;
  $\Delta g \leftarrow \theta\,(g_{\mathrm{new}} - g)$;\quad $g \leftarrow g + \Delta g$\;
  \If{$\mathrm{Im}\,g > 0$}{$g \leftarrow \bar g$\tcp*{project onto physical branch}}
}
$\nu_\eta(\lambda) \leftarrow -(1/\pi)\,\mathrm{Im}\,g$\;
\end{algorithm}

Damping at $\theta = 0.05$ is empirically robust across the tested parameter range. The self-consistent equations have real coefficients and are therefore invariant under simultaneous complex conjugation of $z$, $g$, and $\phi$. At real $z$, solutions come in conjugate pairs, and near the real axis the equations alone do not fix the sign of $\mathrm{Im}\,g$. The iterate is therefore projected onto the physical Stieltjes branch ($\mathrm{Im}\,g < 0$ for $\mathrm{Im}\,z > 0$) at every step. To trace $g(z)$ over a $\lambda$-grid at fixed $\eta$, the sweep starts at the middle of the grid with the default initialisation and proceeds outward in both directions, warm-starting each point from its converged neighbour. 

For the outlier condition, Eq.~\eqref{eq:outlier-noisy}, the same solver is run at real $\lambda$ (with negligible $\eta$) and channel $m$ left out of the channel averages, cf. Sec.~\ref{sec:outliers}. The root finder scans a geometric grid in $\lambda$, at each point obtaining the channel-removed pair $(\tilde g, \tilde\phi)$ by Algorithm~\ref{algo:solver}, and bisects every crossing of the resonance product $a\,\tilde\phi$ with the level $1/w_m$, cf. Eq.~\eqref{eq:outlier-noisy}. Several crossings can occur, since the channel-removed bulk has peaks at its top, and the product, as a function of $\lambda$, falls across every gap of the support and rises back through every peak. The crossings must therefore be classified: an isolated root lies in a gap by definition, cf. Sec.~\ref{sec:outliers}, while inside the support the channel-removed solution is complex and the scan follows only the real part of the product, so a crossing there is not a root of the condition. A gap is an interval on which the channel-removed density $\tilde\nu_\eta(\lambda) = -\pi^{-1}\mathrm{Im}\,\tilde g(\lambda + i\eta)$ vanishes as $\eta \to 0$, and the classifier is the $\eta$-scaling of this density at the crossing: $\tilde\nu_\eta$ is proportional to $\eta$ in a gap and saturates at its $\eta \to 0$ value inside the support. Only gap crossings are kept, and the topmost of these is the isolated root that tracks channel $m$. A channel whose condition has no gap crossing has no isolated root and is not an outlier.

\section{Theory versus simulation in the ranked spectrum}
\label{app:rankplots}

The bridge between a density and a ranked spectrum is the counting function
\begin{align}
\mathcal{N}(\lambda) := N \int_\lambda^{\infty} d\lambda'\,\nu_0(\lambda')\,,
\end{align}
which gives the expected number of eigenvalues above $\lambda$ at leading order. An eigenvalue of rank $r$ is plotted at the level where the expected count equals $r - 1/2$. Exactly $r - 1$ eigenvalues lie above it, and its own unit of spectral mass contributes one half. This is the midpoint convention for comparing order statistics with a continuum quantile. The $1/2$ matters most at the lowest ranks, which a log-log plot emphasises. The theory curve of the ranked spectrum is the counting function's inverse
\begin{align}
\lambda_r = \mathcal{N}^{-1}\!\left(r - \frac12\right), \quad r = 1, \dots, N
\label{eq:rank-curve}
\end{align}
(for $q_V \ge 1$ or $\sigma^2 > 0$; otherwise $r$ runs to $\lfloor N q_V \rfloor$).
The counting function covers both regimes of the spectrum. Where the spectrum is a bulk, $\mathcal{N}$ is smooth and Eq.~\eqref{eq:rank-curve} returns its quantiles. Across an outlier, an isolated peak of spectral mass $1/N$, the counting function steps by one, so Eq.~\eqref{eq:rank-curve} places the top ranks at the outlier positions of Eq.~\eqref{eq:outlier-noisy}. In the noiseless case at $q_V < 1$ the ranks beyond $N q_V$ fall on the Dirac delta at zero of Sec.~\ref{sec:q0-baseline}, and the inverse is taken over the nonzero spectrum. Nonzero noise lifts the delta and all $N$ ranks are finite.

Figure~\ref{fig:ranked-overlays} applies this construction to the ensembles of the density comparisons of Sec.~\ref{sec:checks}. The points are the simulated rank-wise means with their 10th--90th percentile ranges over realisations, and the lines are the theory curves of Eq.~\eqref{eq:rank-curve}. The curves follow the simulation at every rank, across the exponent regimes and across the noise levels: along the discrete top, which the ranked view resolves rank by rank, down the bulk, and onto the noise floor, which compresses the head of the noisy spectra from below.

\section{Theory versus simulation across the latent aspect ratio}
\label{app:qv}

\begin{figure*}[t]
\centering
\includegraphics[width=\textwidth]{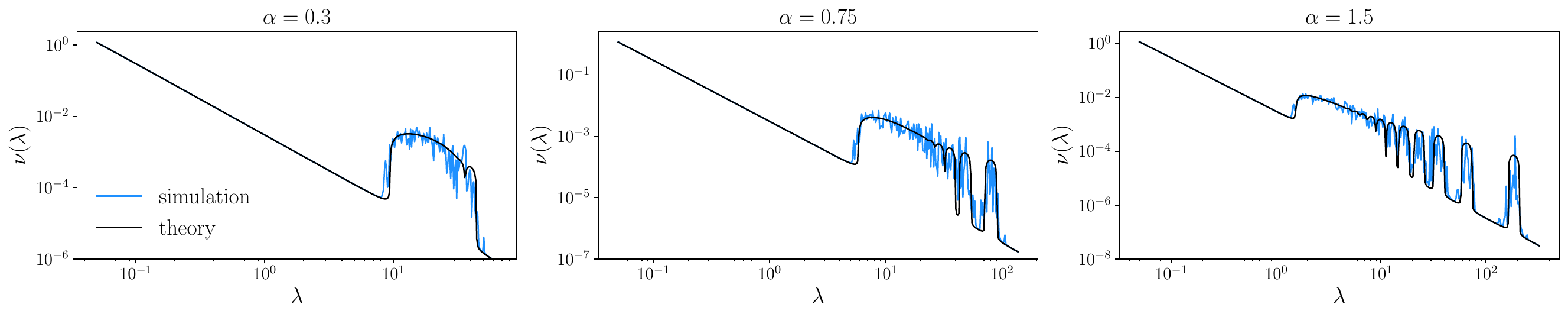}\\
(a) Few latent features ($q_V = 0.05$)\\[8pt]
\includegraphics[width=\textwidth]{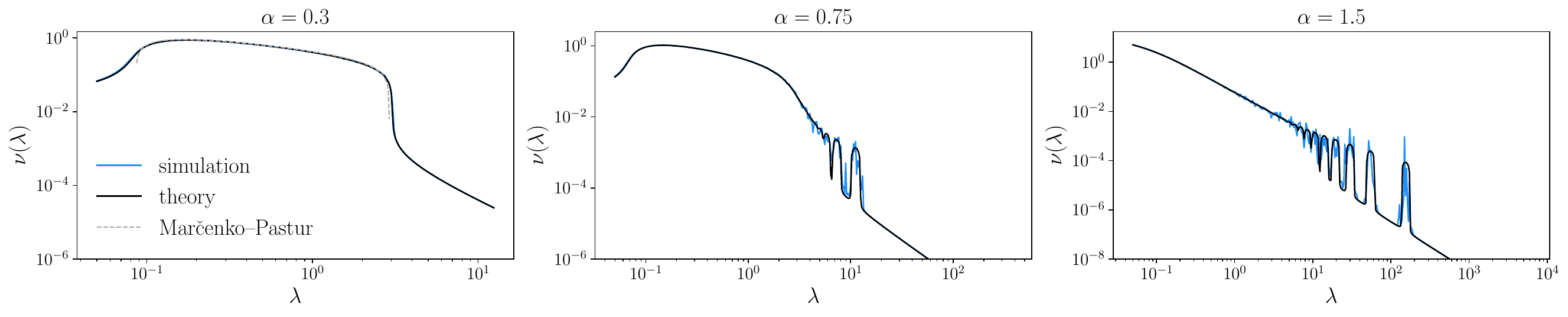}\\
(b) Latent-rich ($q_V = 40$)
\caption{Theory versus simulated densities across the latent aspect ratio. (a) The regime of few latent features, $q_V = 0.05$ ($\ell = 20$). (b) The latent-rich regime, $q_V = 40$ ($\ell = 16000$). Each row spans $\alpha \in \{0.3, 0.75, 1.5\}$. Noiseless ensembles at $N = 400$ and $T = 800$, in the conventions of Fig.~\ref{fig:overlay-alpha}.}
\label{fig:overlay-qv}
\end{figure*}

The density comparisons of Sec.~\ref{sec:checks} are run at latent aspect ratio $q_V = \ell/N = 0.75$ with $N=400$ and $\ell=300$ (and $T=800$). In this appendix we repeat the comparisons for $q_V = 0.05$ and $q_V = 40$, that is at $\ell = 20$ and $\ell = 16000$, respectively, at fixed $N = 400$ and $T = 800$. The latent aspect ratio thus spans three decades from the regime of few latent features ($\ell\ll N$) to the latent-rich regime ($\ell\gg N$). Figure~\ref{fig:overlay-qv} shows the noiseless overlays at $\alpha \in \{0.3, 0.75, 1.5\}$. The theory curves closely follow the simulation in both regimes.

The comparison also shows how the latent aspect ratio shapes the spectrum. First, we note that at super-critical $\alpha$ the positions of the discrete top eigenvalues nearly coincide across $q_V = 0.05$, $0.75$ (Fig.~\ref{fig:overlay-alpha}), and $40$, up to the finite-$\ell$ correction $O(\ell^{1-\alpha})$ to $S_\alpha$, visible only at $\ell = 20$. This is the far-outlier position of Sec.~\ref{sec:outliers} at work. Second, in the noiseless setting, $\gamma^2 = S_\alpha$, and the regimes of the partial sum separate the behaviours:
\begin{align}
\lambda_m \to \dfrac{N\,m^{-\alpha}}{S_\alpha} \simeq
\begin{cases}
\dfrac{(1 - \alpha)\,m^{-\alpha}\ell^{\alpha}}{q_V}\,, & \alpha < 1\,,\\[8pt]
\dfrac{N\,m^{-\alpha}}{\log\ell}\,, & \alpha = 1\,,\\[8pt]
\dfrac{N\,m^{-\alpha}}{\zeta(\alpha)}\,, & \alpha > 1\,,
\end{cases}
\end{align}
cf. Appendix~\ref{app:salpha}.
The latent dimension enters only through the normalisation $S_\alpha$, which converges at super-critical $\alpha$, so the discrete top at fixed $N$ does not move with $q_V$. Sub-critically $S_\alpha$ grows with $\ell$, and the same formula implies a decreasing top as $q_V$ grows. At $q_V = 40$ the $\alpha = 0.75$ hierarchy is compressed relative to $q_V = 0.75$ (Fig.~\ref{fig:overlay-alpha}), its leading position $N/S_\alpha$ falling from about $30$ to about $10$, and at $\alpha = 0.3$ the leading position $N/S_\alpha \approx 0.3$ lies inside the bulk, so the discrete top is absent. In the opposite limit, at $q_V = 0.05$ every exponent carries a strong discrete top, and the $q_V < 1$ spectrum in addition carries the Dirac delta at zero with weight $1 - q_V$ of Sec.~\ref{sec:q0-baseline}.

The sub-critical latent-rich panels approach a classical limit. As $\ell \to \infty$ at fixed $N$ and $T$, each entry of the noiseless data matrix is a sum $X_{tn} = \sum_m U_{tm}V_{mn}$ of independent channel contributions. The central limit theorem applies to it precisely when no single channel retains a finite share of the total variance $S_\alpha$. This is the Lindeberg condition~\cite{lindeberg_1922, billingsley_1995}. Channel $m$ carries the share $m^{-\alpha}/S_\alpha$ of the total variance, largest for the top channel $m = 1$, so the condition reduces to $1/S_\alpha \to 0$ and therefore holds at sub-critical and critical $\alpha$, where $S_\alpha$ diverges. The entries of the data matrix then approach a Gaussian distribution, and the conditional covariance of its columns, $\mathbf{U}\mathbf{\Sigma}\mathbf{U}^T\!/\gamma^2 = \mathbf{Y}/q_U$, concentrates on the identity under the same condition, so the columns become independent isotropic Gaussian vectors and the spectrum collapses onto the Mar\v cenko--Pastur law at aspect ratio $q$. The $\alpha = 0.3$ panel at $q_V = 40$ is close to this collapse. At super-critical $\alpha$ channel $m$ retains the finite variance share $m^{-\alpha}/\zeta(\alpha)$, the condition fails, and the discrete top survives at every $q_V$.

\section{The outlier fluctuation law}
\label{app:fluct}

In this appendix we derive Eq.~\eqref{eq:outlier-fluct}, the Gaussian fluctuation law of an isolated outlier. Gaussian fluctuations of spiked eigenvalues are classical~\cite{bai_yao_2008}, but their scale is model-dependent, already within the deformed Wigner class~\cite{capitaine_donatimartin_feral_2009}, so the coefficient must be derived for our model. We follow the standard approach for low-rank perturbations~\cite{benaych_nadakuditi_2012}: projecting the eigenvector equation onto the directions associated with a latent channel $m$ yields a characteristic equation for the outlier, whose coefficients are quadratic forms in the channel directions. The means of these forms give the deterministic outlier condition, Eq.~\eqref{eq:outlier-noisy}, and their $O(N^{-1/2})$ fluctuations, propagated through the root, give the fluctuation law.

We first assemble the characteristic equation. For this we separate channel $m$ from the data matrix,
\begin{gather}
\frac{\mathbf{X}}{\sqrt{T\gamma^2}} = \tilde{\mathbf{Z}}_\sigma + c\,\hat{u}\,\hat{v}^T\,,\nonumber\\
\hat{u} := \frac{U_{\cdot m}}{\sqrt T}\in\mathbb R^{T\times 1}\,,\quad \hat{v} := \frac{m^{\alpha/2}\,V_{m\cdot}}{\sqrt N}\in \mathbb R^{N\times 1}\,,
\end{gather}
where $\tilde{\mathbf{Z}}_\sigma$ contains the other $\ell - 1$ channels and the noise $\sigma \mathbf{R}/\sqrt{T\gamma^2}$, the vectors $\hat{u}$ and $\hat{v}$ are independent Gaussian vectors with entry variances $1/T$ and $1/N$, respectively, and $c^2 := s\,w_m/q_V$. We define $\tilde{\mathbf{W}} := \tilde{\mathbf{Z}}_\sigma^T\tilde{\mathbf{Z}}_\sigma$, its resolvent $\mathbf{G} := (\lambda - \tilde{\mathbf{W}})^{-1}$, and the direction $\hat{t} := \tilde{\mathbf{Z}}_\sigma^T\hat{u}$. Expanding the normalised Gram matrix gives
\begin{align}
\mathbf{W} &= (\tilde{\mathbf{Z}}_\sigma + c\,\hat{u}\hat{v}^T)^T(\tilde{\mathbf{Z}}_\sigma + c\,\hat{u}\hat{v}^T)\nonumber\\
&= \tilde{\mathbf{W}} + c\,\hat{t}\,\hat{v}^T + c\,\hat{v}\,\hat{t}^T + c^2|\hat{u}|^2\,\hat{v}\,\hat{v}^T\,,
\end{align}
so channel $m$ perturbs $\tilde{\mathbf{W}}$ at rank two, supported on the span of $\hat{v}$ and $\hat{t}$. For an eigenvalue $\lambda$ of $\mathbf{W}$ outside the spectrum of $\tilde{\mathbf{W}}$, the eigenvector equation $(\lambda - \tilde{\mathbf{W}})\,\rho = (\mathbf{W} - \tilde{\mathbf{W}})\,\rho$ inverts to
\begin{align}
\rho = \mathbf{G}\bigl[c\,(\hat{v}^T \rho)\,\hat{t} + \bigl(c\,\hat{t}^T \rho + c^2|\hat{u}|^2\,\hat{v}^T \rho\bigr)\,\hat{v}\bigr]\,.
\end{align}
Projecting this equation onto $\hat{v}$ and $\hat{t}$ gives the homogeneous linear system for the pair $(\hat{v}^T\rho, \hat{t}^T\rho)$,
\begin{align}
\bigl(1 - c\,y - c^2|\hat{u}|^2 p\bigr)\,\hat{v}^T\rho - c\,p\,\hat{t}^T\rho &= 0\,,\nonumber\\
-\bigl(c\,r + c^2|\hat{u}|^2 y\bigr)\,\hat{v}^T\rho + (1 - c\,y)\,\hat{t}^T\rho &= 0\,,
\end{align}
where we have defined
\begin{align}
\label{eq:scalars}
p := \hat{v}^T \mathbf{G}\,\hat{v}\,,\quad r := \hat{t}^T \mathbf{G}\,\hat{t}\,,\quad y := \hat{v}^T \mathbf{G}\,\hat{t}\,,
\end{align}
and $y = \hat{t}^T\mathbf{G}\,\hat{v}$ by symmetry of $\mathbf{G}$.
A nonzero eigenvector requires the determinant of the system to vanish. In the determinant, the $c\,y$ cross terms of the $|\hat{u}|^2$ contribution cancel, leaving the characteristic equation
\begin{align}
\label{eq:char}
J(\lambda) := (1 - c\,y)^2 - c^2\,p\,r - c^2\,|\hat{u}|^2\,p = 0\,.
\end{align}
The root of this equation is a random quantity through the four scalars $p$, $r$, $y$, and $|\hat{u}|^2$. We evaluate it in two steps. First, at the averages of the scalars, which yields the deterministic outlier position, and second, at first order in the fluctuations around the averages, which gives the fluctuation law.

For the first step, we take the averages of the four scalars $p$, $r$, $y$, and $|\hat{u}|^2$ over the channel directions $\hat{u}$ and $\hat{v}$ at fixed $\tilde{\mathbf{Z}}_\sigma$. Since $\hat{u}$ and $\hat{v}$ are independent of $\tilde{\mathbf{Z}}_\sigma$,
\begin{gather}
\mathbb{E}[p] = \frac{1}{N}\,\mathrm{tr}\,\mathbf{G} = \tilde g(\lambda)\,,\quad
\mathbb{E}[y] = 0\,,\quad \mathbb{E}[|\hat{u}|^2] = 1\,,\nonumber\\
\mathbb{E}[r] = \frac{1}{T}\,\mathrm{tr}(\tilde{\mathbf{W}}\,\mathbf{G}) = q(\lambda \tilde g - 1) =: \tilde \varphi - 1\,,
\end{gather}
where $\tilde g$ is the Stieltjes transform of the channel-removed bulk $\tilde{\mathbf{W}}$, the bulk solution introduced in Sec.~\ref{sec:outlier-fluct}. Evaluating Eq.~\eqref{eq:char} at these averages gives
\begin{align}
J = 1 - c^2 \tilde g(\tilde \varphi - 1) - c^2 \tilde g\,,
\end{align}
so the root at the averages satisfies $c^2 \tilde g\tilde \varphi = 1$. Substituting $c^2 = s\,w_m/q_V$ and the identity, Eq.~\eqref{eq:phi-identity-noisy}, for the bulk equations turns this into $s\,\tilde g\,\tilde\phi\,w_m = q_V(1 - q\,n_\sigma\,\tilde g)$, which is the outlier condition, Eq.~\eqref{eq:outlier-noisy}. The root at the averages is thus the deterministic position $\lambda_m$, recovering the condition of Sec.~\ref{sec:outliers} by an independent route.

For the second step, write each scalar quantity entering Eq.~\eqref{eq:char} as its average plus a fluctuation, $q_i = \bar q_i + \delta q_i$ with $q_i \in \{p, r, y, |\hat u|^2\}$ and $\delta q_i = O(N^{-1/2})$. The fluctuations of $\tilde{\mathbf{Z}}_\sigma$ move $\tilde g$ only at $O(1/N)$ and are subleading. We write $\bar J(\lambda) := 1 - c^2\,\tilde g(\lambda)\,\tilde\varphi(\lambda)$ for the characteristic equation evaluated at the averaged scalars. Expanding Eq.~\eqref{eq:char} to first order around the point $\mathcal{P} := (\lambda_m, \bar q_1, \ldots, \bar q_4)$ gives the realised outlier $\hat\lambda_m = \lambda_m + \delta\lambda$ with
\begin{align}
    \delta\lambda = -\frac{\sum_i\,(\partial J/\partial q_i)\big|_{\mathcal{P}}\,\delta q_i}{\bar J'(\lambda_m)}\,,
\end{align}
and the derivatives of $J$ evaluated at $\mathcal{P}$, where they are deterministic:
\begin{align}
\frac{\partial J}{\partial p}\bigg|_{\mathcal{P}} &= -c^2\tilde \varphi\,,\quad
\frac{\partial J}{\partial r}\bigg|_{\mathcal{P}} = \frac{\partial J}{\partial |\hat{u}|^2}\bigg|_{\mathcal{P}} = -c^2 \tilde g\,,\nonumber\\
\frac{\partial J}{\partial y}\bigg|_{\mathcal{P}} &= -2c\,,\quad
\bar J'(\lambda_m) = -c^2\,\partial_\lambda(\tilde g\tilde \varphi)\,.
\end{align}
Each scalar of Eq.~\eqref{eq:scalars} is a quadratic form in the Gaussian vectors $\hat u$ and $\hat v$, with a deterministic kernel for fixed $\tilde{\mathbf{Z}}_\sigma$. Their fluctuations $\delta q_i$ are therefore jointly Gaussian at large $N$, and so is $\delta\lambda$, their linear combination with deterministic coefficients.

The law we are after is its variance, $\mathrm{Var}(\hat\lambda_m) = \mathrm{Var}(\delta\lambda)$, assembled from the covariances of the $\delta q_i$ and the squared derivatives. For Gaussian quadratic forms, Wick contractions give the standard identities
\begin{gather}
\mathrm{Var}(\hat{v}^T\mathbf{M}\hat{v}) = \frac{2\,\mathrm{tr}(\mathbf{M}^2)}{N^2}\,,\quad
\mathrm{Var}(\hat{u}^T\mathbf{M}\hat{u}) = \frac{2\,\mathrm{tr}(\mathbf{M}^2)}{T^2}\,,\nonumber\\
\mathrm{Var}(\hat{u}^T\mathbf{M}\hat{v}) = \frac{\mathrm{tr}(\mathbf{M}\mathbf{M}^T)}{T N}\,,\nonumber\\
\mathrm{Cov}(|\hat{u}|^2, \hat{u}^T\mathbf{M}\hat{u}) = \frac{2\,\mathrm{tr}(\mathbf{M})}{T^2}\,.
\end{gather}
The scalars are of these forms, with kernels $\mathbf{M} = \mathbf{G}$ for $p$, $\mathbf{M} = \tilde{\mathbf{Z}}_\sigma\mathbf{G}$ for $y$, $\mathbf{M} = \tilde{\mathbf{Z}}_\sigma\mathbf{G}\tilde{\mathbf{Z}}_\sigma^T$ for $r$ and its covariance with $|\hat u|^2$, and $\mathbf{M} = \mathbf{I}_T$ for $|\hat u|^2$.
All other covariances among $p$, $r$, $y$, and $|\hat{u}|^2$ vanish because they are odd in $\hat{u}$ or $\hat{v}$ or factorise by the independence of $\hat{u}$ and $\hat{v}$. The required traces can be expressed in terms of two spectral sums of the bulk, $\tilde g=N^{-1}\mathrm{tr}\,\mathbf{G}$ and $ \tau_2:=N^{-1}\mathrm{tr}\,\mathbf{G}^2 = -\tilde g'$: the resolvent identity $\tilde{\mathbf{W}}\mathbf{G} = \lambda \mathbf{G} - \mathbf{I}$ removes every factor $\tilde{\mathbf{W}}$ from the trace of each kernel. For example, $\mathrm{tr}[(\tilde{\mathbf{Z}}_\sigma\mathbf{G}\tilde{\mathbf{Z}}_\sigma^T)^2] = \mathrm{tr}[(\tilde{\mathbf{W}}\mathbf{G})^2]$, which equals $\lambda^2\,\mathrm{tr}\,\mathbf{G}^2 - 2\lambda\,\mathrm{tr}\,\mathbf{G} + N$. We thus obtain the following expressions
\begin{gather}
\mathrm{Var}(p) = \frac{2\tau_2}{N}\,,\quad
\mathrm{Var}(y) = \frac{q\,(\lambda\tau_2 - \tilde g)}{N}\,,\nonumber\\
\mathrm{Var}(r) = \frac{2q^2(\lambda^2\tau_2 - 2\lambda \tilde g + 1)}{N}\,,\quad
\mathrm{Var}(|\hat{u}|^2) = \frac{2q}{N}\,,\nonumber\\
\mathrm{Cov}(r, |\hat{u}|^2) = \frac{2q\,(\tilde \varphi - 1)}{N}\,.
\end{gather}
Collecting the various terms, and using $c^{-2} = \tilde g\tilde \varphi$ to eliminate $c$, gives
\begin{align}
\mathrm{Var}(\hat\lambda_m) = \frac{1}{N}\,\frac{v_m}{\bigl[\partial_\lambda(\tilde g\tilde \varphi)\bigr]^2}\,,
\end{align}
with
\begin{align}
v_m &= 2\,\tilde \varphi^2\,\tau_2 + 2\,q^2 \tilde g^2\bigl(\lambda_m^2\tau_2 - 2\lambda_m \tilde g + 1\bigr)\nonumber\\
&\quad + 4\,q\,\tilde g\tilde \varphi\bigl(\lambda_m\tau_2 - \tilde g\bigr) + 2\,q\,\tilde g^2\bigl(2\tilde \varphi - 1\bigr)\,,
\end{align}
which is the fluctuation law, Eq.~\eqref{eq:outlier-fluct}. The $\tilde \varphi^2\tau_2$ term is the contribution of $p$, the $q^2 \tilde g^2(\cdots)$ term that of $r$, the $4q\tilde g\tilde \varphi(\cdots)$ term that of $y$, and the $2q\tilde g^2(2\tilde \varphi - 1)$ term that of $|\hat{u}|^2$ together with its covariance with $r$.

In the far-outlier limit where $\tilde g \to 1/\lambda$, $\tilde \varphi \to 1$, $\tau_2 \to 1/\lambda^2$, the $y$ and $r$ contributions vanish at leading order and the $p$ and $|\hat{u}|^2$ terms give
\begin{align}
\label{eq:far-outlier-var}
\frac{\mathrm{Var}(\hat\lambda_m)}{\lambda_m^2} \to \frac{2}{N} + \frac{2}{T}\,.
\end{align}
The two terms on the right-hand side are the relative variances of the squared norms $|\hat{v}|^2$ and $|\hat{u}|^2$ of the channel vectors that dominate a far outlier. The $2/T$ term is the classical fixed-population spike variance in this limit~\cite{paul_2007}, while the $2/N$ term is the fluctuation of the channel strength itself, which is frozen in the fixed-population ensembles but random here. Both the full law and the far-outlier limit are verified against simulated outlier ensembles in Fig.~\ref{fig:outlier-check} (Sec.~\ref{sec:outlier-fluct}), including for the noisy case at $\sigma^2 = 2$.

\section{Comparison protocol and the sub-critical outlier check}
\label{app:numerics}

\begin{figure*}[t]
\includegraphics[width=0.49\textwidth]{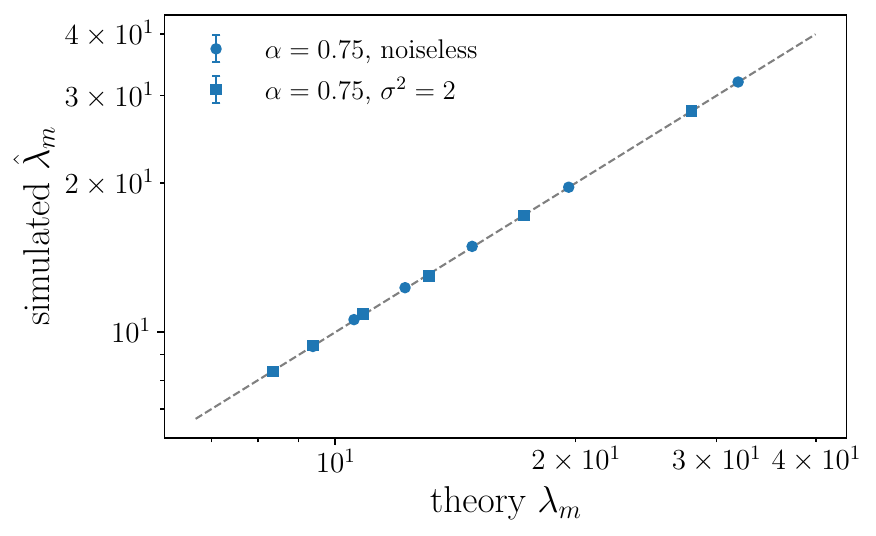}\hfill
\includegraphics[width=0.49\textwidth]{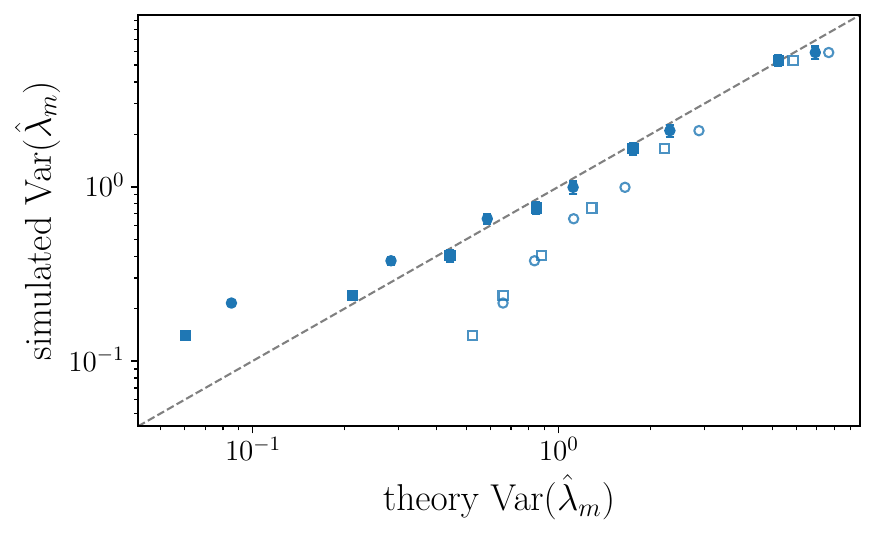}
\caption{Theory versus simulation at the top of the spectrum at sub-critical $\alpha = 0.75$. Positions (left) and fluctuations (right) of the outlier eigenvalues, in the conventions of Fig.~\ref{fig:outlier-check}, for all channel-tracked ranks of the noiseless (circles) and $\sigma^2 = 2$ (squares) ensembles.}
\label{fig:outlier-check-sub}
\end{figure*}

In this appendix we record the protocol behind the outlier comparisons of Sec.~\ref{sec:edge} and their sub-critical companion. For each ensemble we simulate $R = 400$ realisations and compare the rank-wise means and variances of the top eigenvalues against the deterministic positions from the outlier condition, Eq.~\eqref{eq:outlier-noisy}, solved with channel $m$ removed from the bulk, and against the fluctuation law, Eq.~\eqref{eq:outlier-fluct}.

Figure~\ref{fig:outlier-check-sub} repeats the comparison of Fig.~\ref{fig:outlier-check} at sub-critical $\alpha = 0.75$, where the top ranks are only marginally separated. The positions fall on the diagonal at every channel-tracked rank, six per ensemble. The variances follow the fluctuation law through the top four ranks, with the full law measurably closer to the simulation than its far-outlier limit, and the lowest ranks show the same one-sided departure as the super-critical ensembles, with the measured variance exceeding the prediction.


%

\end{document}